\documentclass[a4paper,10pt]{article}
\usepackage{geometry}
\usepackage{physics}
\usepackage{authblk}
\usepackage{graphicx}
\usepackage{caption}
\usepackage{subcaption}
\usepackage{fullminipage}
\usepackage{multirow}
\usepackage{tocloft}
\usepackage{amsmath,amssymb}
\usepackage{slashed}
\usepackage[linkcolor=blue,citecolor=blue,colorlinks=true]{hyperref}
\usepackage{cleveref}
\crefformat{equation}{#2Eq.~(#1)#3}
\Crefformat{equation}{#2Eq.~(#1)#3}
\let\eqref\cref
\usepackage{amssymb}
\usepackage[numbers,sort&compress]{natbib}
\usepackage{mathtools}

\numberwithin{equation}{section}
\DeclareMathAlphabet\mathbfcal{OMS}{cmsy}{b}{n}

\title{Nested warped geometry in a non-flat braneworld scenario}
\author[1]{Arko Bhaumik \thanks{arkobhaumik12@gmail.com}}
\author[2]{Soumitra SenGupta \thanks{tpssg@iacs.res.in}}
\affil[1]{Physics and Applied Mathematics Unit \protect\\ Indian Statistical Institute, Kolkata-700108, India \vspace*{2mm}}
\affil[2]{School of Physical Sciences \protect\\ Indian Association for the Cultivation of Science, Kolkata-700032, India}
\date{}                     
\begin{document}

\maketitle

\begin{abstract}
We generalize nested multiply warped braneworld models by incorporating non-zero brane curvature caused by an effective cosmological constant $\Omega$ induced on the 3-branes. Starting with the doubly warped model, we first analyze the case where the maximally warped brane is identified as the visible brane. For $\Omega<0$, resolution of the gauge hierarchy problem imposes a small upper bound on $|\Omega|$, and can possibly lead to positivity of all the 3-brane tensions. For $\Omega>0$, the latter is not possible but the tuning of the cosmological constant to its tiny observed value is linked to the tuning of the extra dimensional moduli close to the inverse Planck length, justifying the original flat brane approximation. In both regimes, we study the dependence of the scale-clustering of the pair of TeV-branes on the brane cosmological constant and hence its potential role in generating a fermion mass hierarchy between these branes. Identifying the near-maximally warped brane as the visible brane instead opens up regions in the parameter space that allow positive 3-brane tensions for both anti de Sitter and de Sitter branes, subject to non-trivial constraints on the warping parameters. We conclude by generalizing the key results to arbitrary $n$-fold nested warped braneworlds. 
\end{abstract}

\newpage

\tableofcontents

\section{Introduction}
The five dimensional warped braneworld model \cite{rs1} proposed by Randall and Sundrum provides, at first glance, an attractive theoretical framework within which the gauge hierarchy problem of the Standard Model can be naturally resolved without extreme fine tuning of parameters. Equipped with a modulus stabilization mechanism which is either based on BSM fields introduced in the bulk \cite{goldbergerwise1,goldbergerwise2,others1,others2,others3,others4} or purely gravitational in origin \cite{gravstab1,gravstab2,gravstab4,gravstab5}, the low-energy phenomenology of the model predicts a scalar radion typically with TeV-scale mass. The lowest-lying KK-mode of the five-dimensional (5D) graviton is expected to be similarly massive, with enhanced coupling (compared to the massless zero-mode) to SM fields on the visible brane due to the same warping which generates the large mass hierarchy \cite{5dgauge1,5dgauge2,5dgauge3,5dgauge4,5dgauge5,5dgauge6,5dgauge7,5dgauge8,5dgauge9,5dgauge10}. The fact that these theoretical predictions fell within experimentally testable domains accessible to present generation colliders largely motivated the shared interest of both communities in the RS1 model following its conception. However, the non-detection of graviton KK-modes through a variety of channels at the LHC till date \cite{gravexp1,gravexp2,gravexp3,gravexp4,gravexp5,gravexp6} has increasingly constrained the parameter space of the five dimensional paradigm. In particular, a small hierarchy of magnitude $m_H/M_5\sim10^{-2}$ between the fundamental Higgs mass $(m_H)$ and the 5D Planck scale $(M_5)$ is currently required to explain the null results. This so-called ``little hierarchy" implies possible emergence of new physics nearly two orders of magnitude below $M_5$. This is particularly disconcerting since it requires the orbifold radius $r_c$ to be enlarged by a factor of $\mathcal{O}(10^2)$. As the exponential dependence of the warp factor on $r_c$ makes the former sensitive to minor changes in the latter, this possibility becomes severely restricted, which in turn diminishes the efficacy of the 5D model in its originally proposed form. 
\\ \\
Motivated largely by the aforementioned reasons, several higher dimensional generalizations of the original RS model have been proposed, most of which rely upon additional orbifolds of the form $S^1/\mathbb{Z}_2$ \cite{6d1,6d2,6d3,6d4,6d5,6d6,6d7,6d8,6d9,6d10}. In this regard, the possibility of a nested doubly warped flat braneworld model  \cite{doubwarp} is particularly interesting, where the singly warped five dimensional metric from \cite{rs1} is further warped along the direction of a second orbifold, thereby resulting in a ``brane-box" configuration with topology $\left[\mathcal{M}(1,3)\times S^1/\mathbb{Z}_2\right]\times S^1/\mathbb{Z}_2$. The bulk is a slice of $\textrm{AdS}_6$ and the ``walls" are formed by four 4-branes, which intersect at the ``vertices" of the box and give rise to four 3-branes warped to different extents, hence having distinct physical mass scales. In absence of any large hierarchy between the orbifold radii, simultaneous large warping along both directions is forbidden, which causes the four mass scales to be clustered in a near-TeV scale pair and a near-Planck scale pair. Such a setup has certain notable advantages over the 5D scenario. Firstly, the doubly warped structure of the metric renders the first KK mode of the 6D graviton heavier than that of the 5D scenario, while further suppressing its coupling to the SM fields confined to the visible brane \cite{doubgrav}. Unlike the 5D case, these features together help the doubly warped model survive current collider constraints without necessitating any little hierarchy between $m_H$ and $M_6$.  At the same time, significant regions of the parameter space for the doubly warped model are accessible to the LHC and its near-future upgrades and can be probed in upcoming runs, thus marking it as a model of direct experimental interest \cite{doubgrav2}. Secondly, it offers a natural explanation for the mass hierarchy observed among SM fermions if the latter are described by five dimensional fields which are allowed to extend into the bulk. Following standard KK decomposition of such a fermionic wavefunction, it can be shown that boundary kinetic terms localized at the maximally and near-maximally warped 3-branes can alter the effective 4D fermion-scalar Yukawa couplings, thus leading to a splitting among the fermion masses \cite{doubwarp,fermmass}. This splitting is quite small as the physical mass scales of this pair of 3-branes are clustered around $\mathcal{O}$(TeV). Other phenomenological aspects of multiply warped flat braneworld models, e.g. moduli stabilization issues and dynamics of bulk matter and gauge fields, have been studied in \cite{higgsgauge,multscal,gaugematt1,gaugematt2,ArunEtAl,modstab6d}, shedding light on a host of interesting properties. In a recent work \cite{wmass}, it has also been shown how the mass discrepancy of the $W$-boson with the SM prediction as observed by the CDF Collaboration \cite{cdf} can be explained in a doubly warped background.
\\ \\
On the other hand, the negative tension of the TeV-brane has long been an unattractive feature of the 5D RS model, as such branes are known to suffer generically from instability issues in both classical Einstein gravity and beyond \cite{negten1,negten2,negten3,negten4}. Although there exist some dynamical schemes \cite{negtensol1,negtensol2} which make the effective 4D theory on the negative tension brane consistent with general relativity in conjunction with a separate modulus stabilization mechanism (as in \cite{goldbergerwise1}), it would be interesting if the TeV-brane could somehow be imparted positive tension instead and the entire issue avoided. The latter is particularly desirable if one considers string motivated braneworld scenarios, where the corresponding stable D-branes turn out to have positive tensions. Interestingly, this can be achieved for a minimalistic generalization of \cite{rs1} to the case of curved 3-branes as analyzed in \cite{5dcurved1}. The origin of non-zero brane curvature can be traced to an effective 4D cosmological constant ($\Omega$) induced on the 3-branes, leading to either an AdS-Schwarzschild ($\Omega<0$) or a dS-Schwarzschild ($\Omega>0$) geometry. For $\Omega<0$, the tension of the TeV-brane can be rendered positive. Additionally, one obtains a tiny upper bound $(\sim10^{-32})$ on the admissible magnitude of $|\Omega|r_c^2$. Besides successfully averting the issue of the negative tension visible brane, the model is thus interesting from a cosmological viewpoint as well. On the other hand, $\Omega>0$ permits a metastable minimum in the modulus potential whose form is determined solely by the brane vacuum energy, hence obviating the need to invoke any external bulk scalar or to modify the gravitational sector for the purpose of modulus stabilization \cite{gravstab3}. This possibility is arguably more attractive than the conventional bulk field prescription generalized to the curved scenario \cite{5dcurved2}. Some other non-trivial aspects of the 5D curved braneworld, e.g. dynamics of bulk fields and radion-driven inflation, have been studied in \cite{5dcurved3,5dcurvedkk,5dcosm9}.
\\ \\
Motivated by the features of the 5D non-flat and the 6D flat cases separately, it is natural to take the next logical step - generalizing the geometric structure of higher dimensional multiply warped spacetimes (i.e. 6D and beyond) by incorporating non-zero brane curvature. As we show in this work, such models have a variety of interesting properties which set them apart from both individual progenitor models.

\section{The 6d Einstein equations}
Assuming matter fields to be absent, the total gravitational action comprised of the bulk $(S_6)$ and 4-brane $(S_5)$ contributions is $S=S_6+S_5$, where $S_6$ is the Einstein-Hilbert action in presence of the bulk cosmological constant $\Lambda_6$, and $S_5$ contains the brane-localized terms arising out of intrinsic vacuum energy densities of the 4-branes, i.e., the brane tensions.
\begin{equation}
S_6=\int d^4x\int_{-\pi}^{+\pi}dy\int_{-\pi}^{+\pi}dz\sqrt{-g_6}\left(2 M^4\mathbfcal{R}-\Lambda_6\right)
\end{equation}
\begin{equation}
\begin{split}
S_5= & -\int d^4x\int_{-\pi}^{+\pi}dy\int_{-\pi}^{+\pi}dz\sqrt{-g_5}\left[V_1(z)\delta(y)+V_2(z)\delta(y-\pi)\right] \\
& - \int d^4x\int_{-\pi}^{+\pi}dy\int_{-\pi}^{+\pi}dz\sqrt{-\bar g_5}\left[V_3(y)\delta(z)+V_4(y)\delta(z-\pi)\right]
\end{split}
\end{equation}
where $\mathbfcal{R}$ is the six-dimensional Ricci scalar, $M$ is the fundamental (six-dimensional) Planck mass, and $g_5$ and $\bar{g}_5$ are the induced metrics on the corresponding 4-branes. To solve the resulting Einstein equations, we first assume a nested metric ansatz of the following form.
\begin{equation} \label{nest}
ds^2=b(z)^2\left[a(y)^2g_{\mu\nu}dx^\mu dx^\nu+R_y^2dy^2\right]+r_z^2dz^2
\end{equation}
where $R_y$ and $r_z$ are the (constant) radii of the orbifolds. This ansatz differs from the familiar flat brane case in the consideration of an arbitrary $g_{\mu\nu}$ (instead of $\eta_{\mu\nu}$) on the 3-branes. Plugging this ansatz into the total action and extremizing it leads to the following set of Einstein equations. \\ \\
\underline{$\mu\nu$-component}:
\begin{equation} \label{munu}
\begin{split}
& G_{\mu\nu}+g_{\mu\nu}\left(\dfrac{3aa''}{R_y^2}+\dfrac{3a'^2}{R_y^2}+\dfrac{4a^2b\ddot{b}}{r_z^2}+\dfrac{6a^2\dot{b}^2}{r_z^2}\right) \\
& = -\dfrac{\Lambda_6}{4M^4}a^2b^2g_{\mu\nu}-\dfrac{a^2b}{4M^4}g_{\mu\nu}\left[\dfrac{V_1(z)}{R_y}\delta(y)+\dfrac{V_2(z)}{R_y}\delta(y-\pi)+\dfrac{bV_3(y)}{r_z}\delta(z)+\dfrac{bV_4(y)}{r_z}\delta(z-\pi)\right]
\end{split}
\end{equation}
\\
\underline{$yy$-component}:
\\
\begin{footnotesize}
\begin{equation} \label{yy}
-2 M^4\left(R_y^2r_z^2\right)\mathcal{R}+8 M^4\left(3r_z^2a'^2+3R_y^2a^2\dot{b}^2+2R_y^2a^2b\ddot{b}\right)
=-a^2b^2R_y^2r_z\left[r_z\Lambda_6+V_3(y)\delta(z)+V_4(y)\delta(z-\pi)\right]
\end{equation}
\end{footnotesize}
\\
\underline{$zz$-component}:
\\
\begin{footnotesize}
\begin{equation} \label{zz}
-2 M^4\left(R_y^2r_z^2\right)\mathcal{R}+8 M^4\left(3r_z^2a'^2+5a^2\dot{b}^2R_y^2+2r_z^2aa''\right)
=-a^2bR_yr_z^2\left[R_yb\Lambda_6+V_1(z)\delta(y)+V_2(z)\delta(y-\pi)\right]
\end{equation}
\end{footnotesize}
\\
where $G_{\mu\nu}$ is the four dimensional Einstein tensor, $\mathcal{R}=g^{\mu\nu}R_{\mu\nu}$ is the four dimensional Ricci scalar, and the primes and dots denote derivatives wrt $y$ and $z$ respectively. 

In the bulk away from the turning points, the $\delta$-functions can be safely ignored and the brane tension terms no longer contribute. Dividing both sides of \eqref{munu} by $g_{\mu\nu}$ for any pair of $\mu$ and $\nu$ and rearranging terms, it becomes clear that one side of \eqref{munu} depends on $x^\mu$ alone, while the other side is a function only of the compact coordinates $y$ and $z$. This allows us to separate the variables as
\begin{equation} \label{const1}
G_{\mu\nu}=-\Omega g_{\mu\nu}
\end{equation}
\begin{equation} \label{const2}
\dfrac{3aa''}{R_y^2}+\dfrac{3a'^2}{R_y^2}+\dfrac{4a^2b\ddot{b}}{r_z^2}+\dfrac{6a^2\dot{b}^2}{r_z^2}+\dfrac{\Lambda_6}{4M^4}a^2b^2=\Omega
\end{equation}
where $\Omega$ is a global constant, which, as evident from \eqref{const1}, plays the role of an induced four-dimensional cosmological constant. This equation further implies a constant 4D curvature $\mathcal{R}=4\Omega$, which can be used to eliminate $\mathcal{R}$ from \eqref{yy} and \eqref{zz} in terms of $\Omega$. Plugging this result into \eqref{yy} and using \eqref{const2} to simplify, we obtain
\begin{equation} \label{int}
-\dfrac{6aa''}{R_y^2}-\dfrac{4a^2b\ddot{b}}{r_z^2}-\dfrac{6a^2\dot{b}^2}{r_z^2}=\left(\dfrac{a^2b^2}{4M^4}\right)\Lambda_6
\end{equation}
Putting this back into \eqref{zz} and invoking \eqref{yy} to separate the variables $y$ and $z$ finally gives us the following uncoupled ODEs in $a(y)$ and $b(z)$:
\begin{equation} \label{warp1}
a'^2-aa''=\dfrac{\Omega R_y^2}{3}\:\:\:;\:\:\:a''-\alpha^2a=0
\end{equation}
\begin{equation} \label{warp2}
\dot{b}^2-b\ddot{b}=-\dfrac{\alpha^2r_z^2}{R_y^2}
\end{equation} 
where $\alpha$ is a dimensionless constant of separation. Solving these ODEs allows us to determine the two warp factors explicitly, followed by calculations of the brane tensions by incorporating the boundary terms in \eqref{yy} and \eqref{zz}. But the natures of the $a(y)$ and $b(z)$ solutions depend crucially on the positivity (dS-Schwarzschild 3-branes)/negativity (AdS-Schwarzschild 3-branes) of $\Omega$. In the following sections, we focus on the two possibilities separately. 

\section{AdS-Schwarzschild 3-branes ($\Omega<0$)} \label{negom}
In this regime, the general solutions of \eqref{warp1} and \eqref{warp2} are given by
\begin{equation} \label{asol1}
a(y)=\omega_1\textrm{cosh}\left(\textrm{ln}\dfrac{\omega_1}{c_1}+\alpha y\right)\:,\:\:\textrm{with}\:\:\omega_1^2=-\dfrac{\Omega R_y^2}{3\alpha^2}
\end{equation}
\begin{equation} \label{bsol1}
b(z)=\bar{\omega}_1\dfrac{\textrm{cosh}\left(\textrm{ln}\frac{\bar{\omega}_1}{\bar{c}_1}+\beta z\right)}{\textrm{cosh}\left(\textrm{ln}\frac{\bar{\omega}_1}{\bar{c}_1}+\beta\pi\right)}\:,\:\:\textrm{with}\:\:\bar{\omega}_1^2=\dfrac{r_z^2\alpha^2}{R_y^2\beta^2}\textrm{cosh}^2\left(\textrm{ln}\dfrac{\bar{\omega}_1}{\bar{c}_1}+\beta\pi\right)
\end{equation}

As in the flat brane case, we must choose an appropriate normalization for the warp factors such that their values are bounded by $(0,1]$. It is straightforward to see that $a(0)=b(\pi)=1$ is the most natural choice. For $b(\pi)=1$, one obtains $\bar{\omega}_1=1$, and $\bar{c}_1$ becomes a superfluous constant which can be set to unity to simplify \eqref{bsol1} to the following form. 
\begin{equation} \label{bsol11}
b(z)=\dfrac{\textrm{cosh}(\beta z)}{\textrm{cosh}(\beta\pi)}\:,\:\textrm{with}\:\:\:\dfrac{r_z\alpha}{R_y\beta}\textrm{cosh}(\beta\pi)=1
\end{equation} 
Apparently, the solution of $b(z)$ is identical to the solution in the flat brane limit (i.e., with $\Omega\to0$), and the relation between the moduli $\alpha$ and $\beta$ is exactly identical to the equality which relates their flat limit counterparts $c$ and $k$. In hindsight, this makes sense because of the structure of the metric in \eqref{nest}. Due to nested warping, the information about the curvature of the 3-branes is encapsulated principally in $a(y)$, and affects $b(z)$ implicitly through the relation between $\alpha$ and $\beta$. Next, one can evaluate $c_1$ from $a(0)=1$ as
\begin{equation}
a(0)=\omega_1\textrm{cosh}\left(\textrm{ln}\dfrac{\omega_1}{c_1}\right)=1\quad\implies\quad c_1=1+\sqrt{1-\omega_1^2}
\end{equation}
where we have taken only the positive discriminant solution, as in the limit $\omega_1\to0$ one needs to ensure $c_1\to2$ in order to reduce \eqref{asol1} to the well-known flat result $a(y)\to e^{-\alpha y}$. The admissible range of $\omega_1^2$ is clearly $\omega_1^2\in\left[0,1\right]$. As the final step, the solutions $a(y)$ and $b(z)$ need to be plugged in \eqref{const2} so that $\beta$ can be obtained in terms of the fundamental parameters:
\begin{equation} \label{bsol12}
\beta=r_z\sqrt{-\dfrac{\Lambda_6}{40M^4}}
\end{equation}
which, like \eqref{bsol11}, expectedly mimicks the flat brane result and requires $\Lambda_6<0$ to be physically meaningful, i.e., the bulk should be AdS$_6$. 

We can now substitute the warp factors directly in \eqref{munu}, use \eqref{const1} and \eqref{const2} to eliminate the bulk part, and integrate over infinitesimal $\varepsilon$-intervals across each boundary point to obtain the corresponding brane tension. Owing to $\mathbb{Z}_2$ symmetry of the orbifolds, one must take care to replace $y$ with $|y|$ in the expression of $a(y)$ once the boundaries are taken into account. 
\begin{equation} \label{V11}
V_1(z)=24M^2\sqrt{-\dfrac{\Lambda_6}{40}\left(1-\omega_1^2\right)}\:\textrm{sech}(\beta z)
\end{equation}
\begin{equation} \label{V12}
V_2(z)=24M^2\sqrt{-\dfrac{\Lambda_6}{40}}\:\textrm{tanh}\left(\textrm{ln}\dfrac{\omega_1}{c_1}+\alpha\pi\right)\textrm{sech}(\beta z)
\end{equation}
\begin{equation} \label{V13}
V_3(y)=0
\end{equation}
\begin{equation} \label{V14}
V_4(y)=32M^2\sqrt{-\dfrac{\Lambda_6}{40}}\:\textrm{tanh}(\beta\pi)
\end{equation}
where we have exploited the normalization $a(0)=1$ (alongside the fact $\omega_1<c_1$) and the $\alpha-\beta$ relation from \eqref{bsol11} to simplify. The explicitly coordinate-dependent term of each brane tension is essentially the same as that of the flat case. Only the constant coefficients are modified due to the presence of non-zero brane curvature. It is obvious that in the limit $\omega_1\to0$, the entire set \eqref{V11}$-$\eqref{V14} reduces to the results derived in \cite{doubwarp} for the flat doubly warped model. 

Owing to the relation between $\alpha$ and $\beta$ in \eqref{bsol11}, it can be argued that $\alpha$ and $\beta$ cannot both be large if, based on naturalness, one assumes $R_y\sim r_z$, i.e., there is no considerable hierarchy between the two extra dimensional moduli. Instead, one must have either $\alpha>\beta$ or $\alpha<<\beta$, which respectively cause warping predominantly along either $y$ or $z$ direction. The maximally warped 3-brane is the $(y=\pi,z=0)$ brane, which we identify as the visible brane for now, based on the conservative assumption that no other brane has a physical mass scale lower than that of the visible brane. The total warp factor is given by $a(\pi)b(0)$, which can be factorized as
\begin{equation} \label{warping}
a(\pi)=\omega_1\textrm{cosh}\left(\textrm{ln}\dfrac{\omega_1}{c_1}+x_1\right)=10^{-n_1}\:,\:\:b(0)=\textrm{sech}(x_2)=10^{-n_2}
\end{equation}
where $x_1=\alpha\pi$ and $x_2=\beta\pi$, and $n_1$ and $n_2$ are positive constants quantifying the extents of warping along the $y$ and $z$ direction respectively. For successful resolution of the gauge hierarchy problem, one requires $n_1+n_2\simeq16$. We show that in absence of a large hierarchy between $R_y$ and $r_z$, $n_1$ and $n_2$ cannot be of similar magnitudes. The exact solutions of \eqref{warping} are
\begin{equation} \label{exsol1}
e^{-x_1}=\dfrac{10^{-n_1}}{c_1}\left(1\pm\sqrt{1-\omega_1^210^{2n_1}}\right)
\end{equation}
\begin{equation} \label{exsol2}
e^{x_2}=10^{n_2}\left(1\pm\sqrt{1-10^{-2n_2}}\right)
\end{equation}
from which one immediately obtains the constraint $\omega_1^2\leq10^{-2n_1}$. Now, if possible, let $n_1$ and $n_2$ both be large, e.g. $n_1\sim n_2\sim8$. Then $\omega_1^2$ is extremely small and renders $c_1\approx2$. We can write $\omega_1^2=10^{-N}$, with the minimum allowed value of $N$ being $N_{min}=2n_1$. For $N\to\infty$, the solution of \eqref{exsol1} reduces to the familiar flat result $x_1=n_1\textrm{ln}10$, whereas for $N=N_{min}=2n_1$ we have $x_1\approx n_1\textrm{ln}10+\textrm{ln}2$, i.e., the flat result plus a minor correction due to the brane curvature. However, for $N>>N_{min}$, one obtains \textit{two} distinct values of $x_1$ corresponding to the two roots of \eqref{exsol1}, both of which are physically valid solutions
\begin{equation} \label{exsolexp1}
x_1^{(1)}\:\approx\:n_1\textrm{ln}10+\dfrac{1}{4}\:10^{-(N-2n_1)}\:\:,\:\:x_1^{(2)}\:\approx\:(N-n_1)\textrm{ln}10+\textrm{ln}4
\end{equation}
where $x_1^{(1)}$ is practically the flat result, and $x_1^{(2)}$ is slightly larger than $x_1^{(1)}$. Hence, in presence of brane curvature, a given degree of large warping along $y$ direction can owe its origin to two distinct values of $\alpha$. Being of similar magnitudes, these two values do not give rise to any new hierarchy that we need to worry about. On the other hand, \eqref{exsol2} admits only one solution for large $n_2$:
\begin{equation} \label{exsolexp2}
x_2\:\approx\:n_2\textrm{ln}10+\textrm{ln}2
\end{equation}
For close values of $n_1$ and $n_2$ (e.g. $n_1\sim n_2\sim8$), it is clear that $\alpha$ and $\beta$ must be of similar magnitudes (e.g. $\alpha\sim\beta\sim6$), which leads back to the large $R_y/r_z$ hierarchy that we are trying to avoid. The only way out is to consider dominant warping along \textit{either} $y$ \textit{or} $z$, corresponding to either $\alpha>\beta$ or $\alpha<<\beta$ as argued in the following sections.

\subsection{Dominant warping along $y$ ($\alpha>\beta$)} \label{negomy}
It suffices to consider $\alpha\sim10$ and $\beta\sim0.1$ to avoid a significantly large $R_y/r_z$ ratio. These values correspond roughly to $n_1\sim15$ and $n_2\sim1$. The largeness of $n_1$ makes the two solutions of $\alpha$ given in \eqref{exsolexp1} valid in this regime. As $n_2$ is not quite as large, the approximate solution from \eqref{exsolexp2} does not hold as good, and one needs to consider the exact version in \eqref{exsol2} instead for a precise value of $\beta$ corresponding to a given $n_2$.

Thus, for $\Omega<0$ and dominant warping along the direction of the first orbifold, the gauge hierarchy problem can be resolved for two distinct values of the orbifold modulus, one of which ($x_1^{(1)}$) is very close to the value obtained in the flat brane scenario and the other one ($x_1^{(2)}$) slightly larger. As noted earlier, $n_1\sim15$ puts a very small upper bound on the induced cosmological constant by ensuring $\omega_1^2\lesssim10^{-30}$. For $\alpha\sim10$, \eqref{asol1} implies that this bound corresponds to $|\Omega| R_y^2\lesssim10^{-28}$. 

\begin{figure}[h] 
\centering
\begin{subfigure}[h]{.46\textwidth}
\begin{center}
\includegraphics[width=\linewidth, height=0.25\textheight]{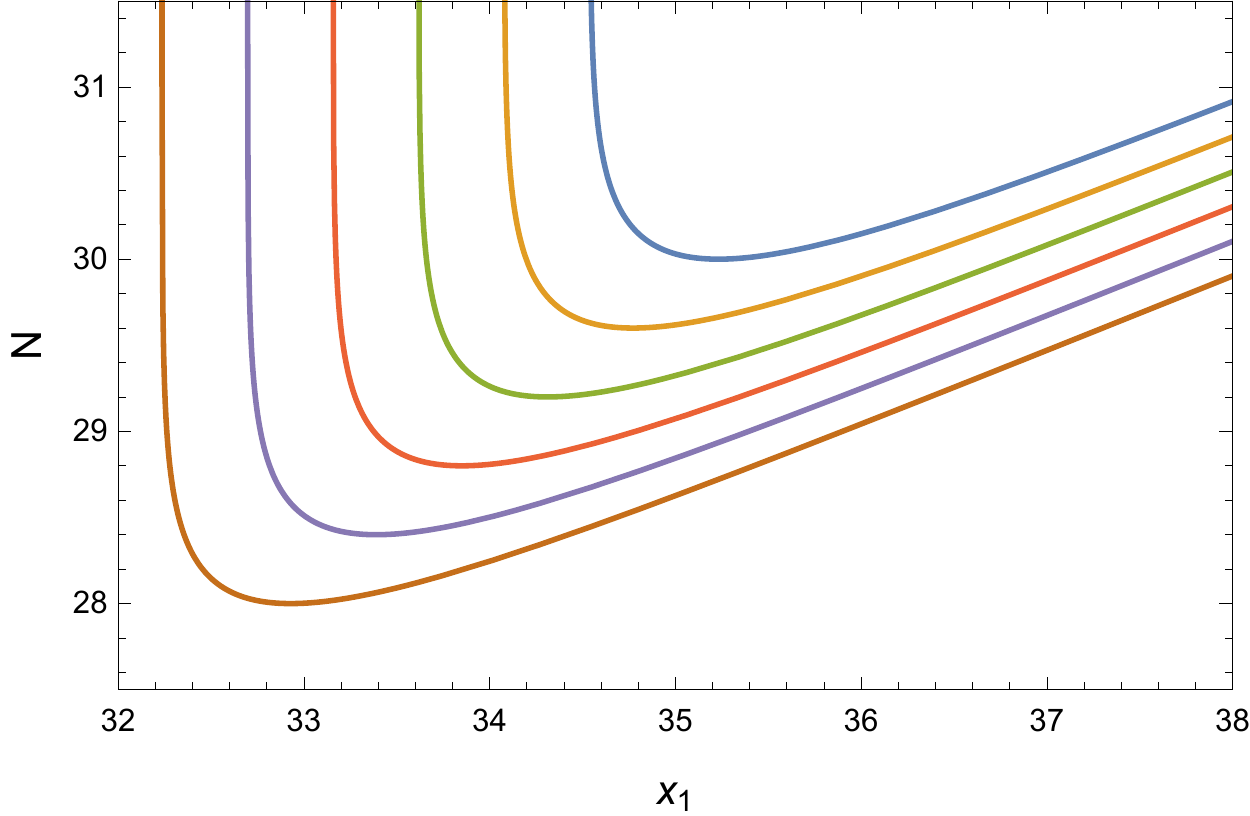}
\caption{\:}
\label{1a}
\end{center}
\end{subfigure}
\hfill
\begin{subfigure}[h]{.53\textwidth}
\begin{center}
\includegraphics[width=\linewidth, height=0.25\textheight]{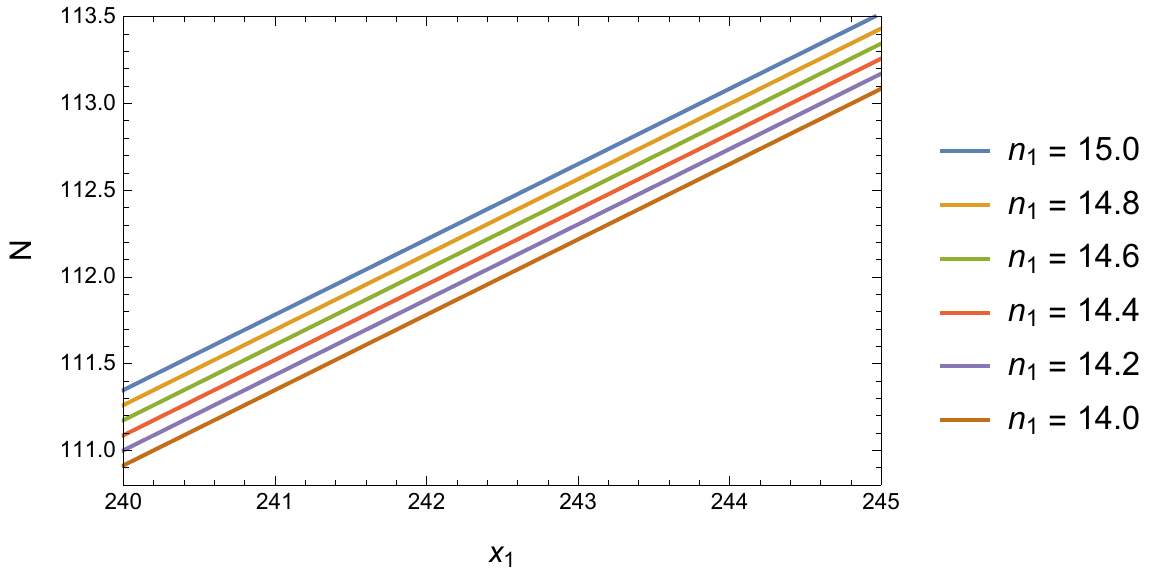}
\caption{\:}
\label{1b}
\end{center}
\end{subfigure}
\caption{\begin{footnotesize} Plots of $N$ versus $x_1$ for a few representative values of $n_1=14-15$, for negative brane cosmological constant and $\alpha>\beta$. Plot (a) shows the behaviour near the global minimum $N_{min}=2n_1$ alongside the divergence at $x_1=n_1\textrm{ln}10$. Plot (b) shows the approximately linear continuation of the curves far from $N_{min}$. \end{footnotesize}}
\label{fig1}
\end{figure}

In Figure \ref{1a}, using the equation $a(\pi)=10^{-n_1}$ from \eqref{warping}, we have plotted $N$ as a function of $x_1$ for a few representative values of $n_1$ which result in no large radius hierarchy. The analytical form of $N(x_1)$ near $N_{min}$ corresponds to the first solution $x_1^{(1)}$ from \eqref{exsolexp1}, which is the dominant solution near the minimum. For each value of $n_1$, $N$ diverges and we obtain $\omega_1^2=0$ at the limiting value of $x_1=n_1\textrm{ln}10$, and a global minimum exists at $N_{min}=2n_1$. The linear relation between $N_{min}$ and $n_1$ is also evident from the set of curves shown in the plots. Figure \ref{1b} shows the behaviour far from the minimum, i.e., for $N>>N_{min}$, where the second solution $x_1^{(2)}$ from \eqref{exsolexp1} dominates and makes $N(x_1)$ approximately linear.   

Of particular interest in this regime are the 3-brane tensions, which are given by pairwise algebraic sums of the 4-brane tensions at the corresponding boundary points. As $\omega_1<c_1$, the forms of \eqref{V11}$-$\eqref{V14} imply that all the 4-brane tensions are decidedly non-negative except $V_2(z)$, which is difficult to tell at first glance. If $V_2(z)$ is positive, all the corner branes have positive tensions, which is a satisfactory feature as negative tension branes are often intrinsically unstable. The crucial sign-determining term of $V_2(z)$ is
\begin{equation} \label{tanh}
\textrm{tanh}\left(\textrm{ln}\dfrac{\omega_1}{c_1}+\alpha\pi\right)\:\approx\:\dfrac{1-\frac{4}{\omega_1^2}e^{-2x_1}}{1+\frac{4}{\omega_1^2}e^{-2x_1}}
\end{equation}
which tends to $-1$ for $x_1=x_1^{(1)}$, and to $+1$ for $x_1=x_1^{(2)}$. Consequently, one ends up with two possible tensions of the $y=\pi$ brane, corresponding to the two allowed values of the modulus $x_1$ that solve the hierarchy problem:
\begin{equation}
V_2^{(1)}(z)\:\approx\:-24M^2\sqrt{-\dfrac{\Lambda_6}{40}}\:\textrm{sech}(kz)\:,\:\:\textrm{for}\:\: x_1=x_1^{(1)}=n_1\textrm{ln}10+\dfrac{1}{4}\:10^{-(N-2n_1)}
\end{equation}
\begin{equation}
V_2^{(2)}(z)\:\approx\:+\:24M^2\sqrt{-\dfrac{\Lambda_6}{40}}\:\textrm{sech}(kz)\:,\:\:\textrm{for}\:\: x_1=x_1^{(2)}=(N-n_1)\textrm{ln}10+\textrm{ln}4
\end{equation}
On the other hand, the tension $V_1(z)$ of the $y=0$ brane is independent of $x_1$, and is approximately equal to $V_2^{(2)}(z)$ for small $\omega_1$. For $x_1=x_1^{(2)}$, all the 4-brane tensions are thus positive, with the sole exception of $V_3=0$. The tensions of the four 3-branes at the corners are therefore strictly positive. Importantly, if we identify the maximally warped $(y=\pi,z=0)$ brane as the visible brane, then its tension is simply equal to $V_2(0)$. This demonstrates that the second solution $x_1^{(2)}$, which arises solely in the context of non-flat geometry, is of key importance in giving us a positive tension visible brane - a feature impossible in the flat case. 
\\

\subsection{Dominant warping along $z$ ($\alpha<<\beta$)} \label{negomz}
In the opposite regime, we have $\beta\gtrsim10$, which implies an extremely tiny upper bound of say $\alpha\sim10^{-12}$ so as to eliminate the unwanted $R_y/r_z$ hierarchy. Thus, the contribution of $\alpha$ becomes negligibly small in the warp factor $a(\pi)$, which approaches unity and corresponds to an infinitesimally small $n_1<<1$. Hence, warping occurs predominantly along $z$, with the extent of shared warping between the two directions being many orders of magnitude smaller than the $\alpha>\beta$ case. 

\begin{figure}[h]
\centering
\includegraphics[width=0.6\textwidth,height=0.4\textwidth]{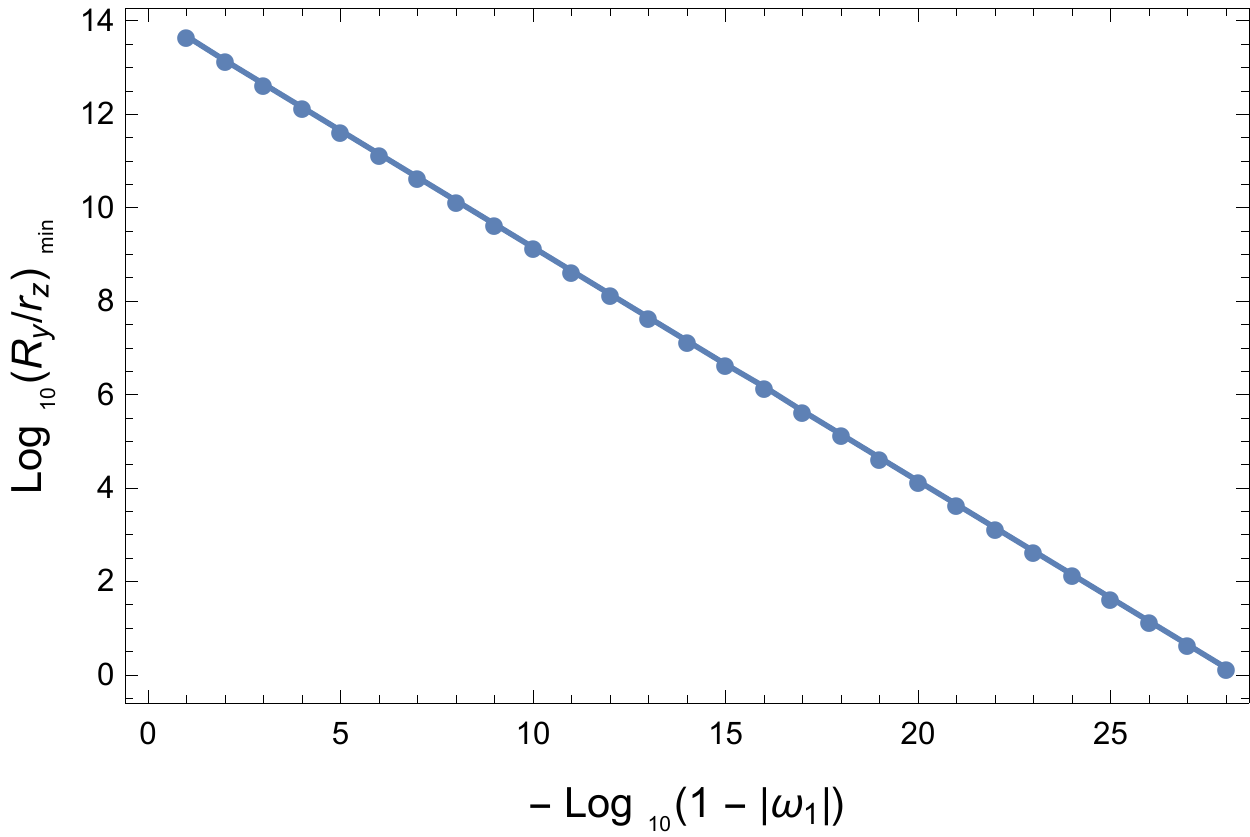}
\caption{\begin{footnotesize} For negative brane cosmological constant and $\alpha<<\beta$, logarithmic plot of the magnitude of the minimum admissible radius ratio $(R_y/r_z)_{min}$ consistent with positive tension of the $y=\pi$ brane, versus the required degree of fine tuning of $\omega_1^2$ close to unity. \end{footnotesize}}
\label{tuning}
\end{figure}

This regime is less interesting than the previous one, as it effectively reduces to a singly warped six-dimensional geometry with a nearly unwarped five-dimensional submanifold. However, it still allows the possibility that all 3-brane tensions can be positive. As before, the determining feature is the positivity of the hyperbolic term in $V_2(z)$, for which we require
\begin{equation} \label{tenpos}
\textrm{ln}\dfrac{\omega_1}{c_1}+\alpha\pi\geq0\quad\implies\quad\left(\dfrac{R_y}{r_z}\right)\dfrac{\beta\pi}{\textrm{cosh}(\beta\pi)}\geq\textrm{ln}\dfrac{c_1}{\omega_1}
\end{equation}
where we have used \eqref{bsol11} to arrive at the second inequality. As the contribution of $\alpha$ is negligible, we need $\beta\simeq12$ for an overall warping of order $10^{-17}$ which can resolve the hierarchy problem. This leads to the following minimum value of the $R_y/r_z$ ratio compatible with the condition that $V_2(z)$ is positive.
\begin{equation} \label{ratmin}
\left(\dfrac{R_y}{r_z}\right)_{min}\sim\:\textrm{ln}\left(\dfrac{1+\sqrt{1-\omega_1^2}}{\omega_1}\right)\times10^{14}
\end{equation}
First, note that a small value of $n_1$ allows $\omega_1$ to be correspondingly close to unity according to the constraint placed by \eqref{exsol1}. Conversely, for a negligibly small $\alpha$, one can treat $\omega_1^2$ as a free parameter and tune it sufficiently close to $1$ to generate a very small warping $n_1$ along $y$. According to \eqref{tenpos} and \eqref{ratmin}, such a fine tuning can render $V_2(z)$ positive while simultaneously ensuring no large hierarchy comes into play between $R_y$ and $r_z$. In Figure \ref{tuning}, we have shown the roughly linear dependence of $(R_y/r_z)_{min}$ on the tuning of $|\omega_1|$ close to unity. As an example, for $(R_y/r_z)_{min}$ to be no larger than $10^2$, this tuning has to be very delicate with $(1-|\omega_1|)\lesssim10^{-23}$. This reintroduces the fine tuning problem in a new guise, and renders the prospect of having a positive tension $(\pi,0)$ brane in this regime problematic, albeit theoretically possible. Note that although $\omega_1^2$ needs to be tuned very close to $1$, the smallness of $\alpha$ nonetheless manages to keep $\Omega R_y^2$ bounded above by a very small value. For example, $\alpha\sim10^{-14}$ roughly gives $|\Omega| R_y^2\lesssim10^{-28}$, which is similar to the upper bound obtained earlier for $\alpha>\beta$. 

In the small $\alpha$ regime, the curved braneworld scenario exhibits another interesting property. Due to unequal warpings along $y$ and $z$, a salient feature of the doubly warped model (both curved and flat) is the clustering of pairs of 3-branes around the fundamental scale and the maximally warped scale, e.g. for $\alpha<<\beta$, the $(0,\pi)$ and $(\pi,\pi)$ branes cluster around the fundamental scale $M$, while the $(0,0)$ and $(\pi,0)$ branes cluster around the maximally warped (TeV) scale (see \cite{doubwarp}). As the effect of the induced cosmological constant shows up only through the subdominant warp factor $a(y)$, we expect $\Omega$ to directly affect the degree of clustering between the mass scales of the $(0,0)$ and $(\pi,0)$ branes. The total warp factors are $a(0)b(0)$ and $a(\pi)b(0)$ respectively. Since $a(0)=1$, the ratio of the maximally and near-maximally warped scales is simply equal to $a(\pi)$. Owing to the smallness of $\alpha$, we can expand $a(\pi)$ up to first order in $\alpha$ to obtain its dependence on $\omega_1^2$.
\begin{equation}
\dfrac{\textrm{Mass scale of maximally warped}\:(\pi,0)\:\textrm{brane}}{\textrm{Mass scale of near-maximally warped}\:(0,0)\:\textrm{brane}}=a(\pi)\:\approx\:1-\alpha\pi\sqrt{1-\omega_1^2}
\end{equation}
For all $0<\omega_1^2\leq1$, this ratio is larger than $(1-\alpha\pi)$ as obtained for the flat case with $\omega_1=0$. In other words, the scales of the TeV-branes are clustered more closely when they are negatively curved than when they are flat. For $\alpha\sim10^{-14}$, the ratio virtually tends to unity for the aforementioned tuning of $\omega_1^2$ close to $1$ which is necessary for a positive tension $(\pi,0)$ brane. If the latter condition is relaxed, the ratio can be comparatively smaller than unity. For all the cases though, deviations from the flat limit are imperceptibly small from a practical point of view, as the smallness of $\alpha$ allows any difference to show up only at the 14th decimal place or beyond.

It should be emphasized, however, that these results depend crucially on how stringently we wish to avoid a hierarchy between $R_y$ and $r_z$, which, in turn, dictates how small an $\alpha$ needs to be chosen. For example, if one ceases to bother about the former, one can choose significantly larger values of $\alpha$ instead, leading to larger differences from the corresponding flat results. Of course, such choices are largely not compatible with the naturalness assumption and simply bring back the hierarchy problem in a new avatar, hence are not helpful from a physical point of view. To drive home the point, we tabulate in Table \ref{split1} a few combinations of $\alpha$ and $\omega_1$ which affect TeV scale clustering in a pronounced manner, alongside the orders of corresponding $R_y/r_z$ ratios.
\\
\begin{table}[ht]
\begin{center}
\begin{tabular}{|c|c|c|c|c|}
\hline
$\alpha$ & $a(\pi)$ in flat limit & $\sim R_y/r_z$ & $\omega_1$ & $a(\pi)$ for given $\omega_1$ \\
\hline\hline
\multirow{3}{*}{$10^{-6}$} & \multirow{3}{*}{$0.9999968584$} & \multirow{3}{*}{$\sim10^9$} & $0.3$ & $0.9999970031$ \\ & & & $0.6$ & $0.9999974867$ \\ & & & $0.9$ & $0.9999986306$ \\
\hline
\multirow{3}{*}{$10^{-4}$} & \multirow{3}{*}{$0.9996858901$} & \multirow{3}{*}{$\sim10^{11}$} & $0.3$ & $0.9997003605$ \\ & & & $0.6$ & $0.9997487219$ \\ & & & $0.9$ & $0.9998631105$ \\
\hline
\multirow{3}{*}{$10^{-2}$} & \multirow{3}{*}{$0.9690724263$} & \multirow{3}{*}{$\sim10^{13}$} & $0.3$ & $0.9705197070$ \\ & & & $0.6$ & $0.9753566452$ \\ & & & $0.9$ & $0.9867973832$ \\
\hline
\end{tabular}
\end{center}
\vspace*{-2mm}
\caption{\begin{footnotesize} A few representative combinations of $\alpha$ and $\omega_1$ which are shown to considerably affect the clustering of the TeV scale $(0,0)$ and $(\pi,0)$ branes, as compared to the flat limit result. However, they are all accompanied by dangerously large radii ratios which spoil naturalness and introduce a new hierarchy in the model. For all the combinations, $\beta\simeq12$ has been fixed beforehand to obtain the desired amount of dominant warping along $z$. \end{footnotesize}}
\label{split1}
\end{table}

\section{dS-Schwarzschild 3-branes ($\Omega>0$)} \label{posom}
In this regime, the general solutions of \eqref{warp1} and \eqref{warp2} are given by
\begin{equation} \label{asol2}
a(y)=\omega_2\:\textrm{sinh}\left(\textrm{ln}\dfrac{c_2}{\omega_2}-\alpha y\right)\:,\:\:\textrm{with}\:\:\omega_2^2=\dfrac{\Omega R_y^2}{3\alpha^2}
\end{equation}
\begin{equation} \label{bsol2}
b(z)=\bar{\omega}_2\dfrac{\textrm{cosh}\left(\textrm{ln}\frac{\bar{\omega}_2}{\bar{c}_2}+\beta z\right)}{\textrm{cosh}\left(\textrm{ln}\frac{\bar{\omega}_2}{\bar{c}_2}+\beta\pi\right)}\:,\:\:\textrm{with}\:\:\bar{\omega}_2^2=\dfrac{r_z^2\alpha^2}{R_y^2\beta^2}\textrm{cosh}^2\left(\textrm{ln}\dfrac{\bar{\omega}_2}{\bar{c}_2}+\beta\pi\right)
\end{equation}
As before, the usual normalization of $b(\pi)=1$ yields $\bar{\omega}_2=1$ and renders $\bar{c}_2$ a superfluous constant which can be set to unity. This makes \eqref{bsol11} hold good. Further, normalizing with $a(0)=1$ gives the following solution of $c_2$
\begin{equation} \label{poswarp2}
a(0)=\omega_2\textrm{sinh}\left(\textrm{ln}\dfrac{c_2}{\omega_2}\right)=1\quad\implies\quad c_2=1+\sqrt{1+\omega_2^2}
\end{equation}
where we consider only the solution which allows $c_2\to2$ so that $a(y)\to e^{-\alpha y}$ in the limit $\omega_2\to0$. Unlike the AdS case, \eqref{poswarp2} itself does not require $\omega_2^2$ to be bounded above by unity. However, in order to ensure positivity of the warp factor $a(\pi)$, the relation $\omega_2/c_2<e^{-\alpha\pi}$ needs to hold, furnishing an upper limit for $\omega_2$ corresponding to any given value of $\alpha$. The relation between $\beta$ and $r_z$ remains the same as in \eqref{bsol12}. Evaluating the brane tensions proceeds in the same manner as described earlier and yields:
\begin{equation} \label{V21}
V_1(z)=24M^2\sqrt{-\dfrac{\Lambda_6}{40}\left(1+\omega_2^2\right)}\:\textrm{sech}(\beta z)
\end{equation}
\begin{equation} \label{V22}
V_2(z)=-24M^2\sqrt{-\dfrac{\Lambda_6}{40}}\textrm{coth}\left(\textrm{ln}\dfrac{c_2}{\omega_2}-\alpha\pi\right)\textrm{sech}(\beta z)
\end{equation}
\begin{equation} \label{V23}
V_3(y)=0
\end{equation}
\begin{equation} \label{V24}
V_4(y)=32M^2\sqrt{-\dfrac{\Lambda_6}{40}}\textrm{tanh}(\beta\pi)
\end{equation}
Like in the AdS regime, one cannot have large simultaneous warping along both directions if one precludes a large $R_y/r_z$ hierarchy. This can be shown by setting $a(\pi)=10^{-n_1}$ and $b(0)=10^{-n_2}$ as before, and solving for $\alpha$ and $\beta$ provided $n_1\sim n_2$ and $n_1+n_2\simeq16$. The allowed regimes remain the same, i.e., either $n_1\sim15$ corresponding to $\alpha\sim10$ and $\beta\sim0.1$, or $n_2\:\approx\:16$ with $\beta\sim12$ and $\alpha\sim10^{-16}$. The analysis is not repeated here as it is exactly the same as in the AdS case. Instead, we move directly on to the two possible scenarios.

\subsection{Dominant warping along $y$ ($\alpha>\beta$)} \label{posomy}
The crucial difference with the AdS case is the fact that for $\alpha>\beta$, the large warping $n_1\sim15$ along $y$ can be generated only by a unique value of $\alpha$, unlike the two degenerate solutions found in \eqref{exsolexp1}. This can be seen by simply solving $a(\pi)=10^{-n_1}$ explicitly, which gives the solution
\begin{equation} \label{exsol3}
\dfrac{c_2}{2}\left(e^{-x_1}-\dfrac{\omega_2^2}{c_2^2}e^{x_1}\right)=10^{-n_1}\:\:\implies\:\:e^{-x_1}=\dfrac{10^{-n_1}}{c_2}\left(1+\sqrt{1+\omega_2^210^{2n_1}}\right)
\end{equation}
where $x_1=\alpha\pi$ as defined earlier. The other solution with a negative discriminant renders $e^{-x_1}<0$, hence must be discarded. 

It is straightforward to see that all the brane tensions from \eqref{V21}$-$\eqref{V24} are strictly non-negative, with the sole exception of $V_2(z)$ whose nature depends on the sign of the constant hyperbolic coefficient as follows.
\begin{equation}
\textrm{coth}\left(\textrm{ln}\dfrac{c_2}{\omega_2}-\alpha\pi\right)=\dfrac{1+\frac{\omega_2^2}{c_2^2}e^{2x_1}}{1-\frac{\omega_2^2}{c_2^2}e^{2x_1}}
\end{equation}
The numerator is clearly positive, and so is the denominator by virtue of the positivity of the warp factor and \eqref{exsol3}. This makes the $\textrm{coth}$ term positive, which renders $V_2(z)$ from \eqref{V22} negative. Thus, it is impossible to have a positive tension $(\pi,0)$ brane in this case. 

\begin{figure}[h]
\centering
\includegraphics[width=0.8\textwidth,height=0.5\textwidth]{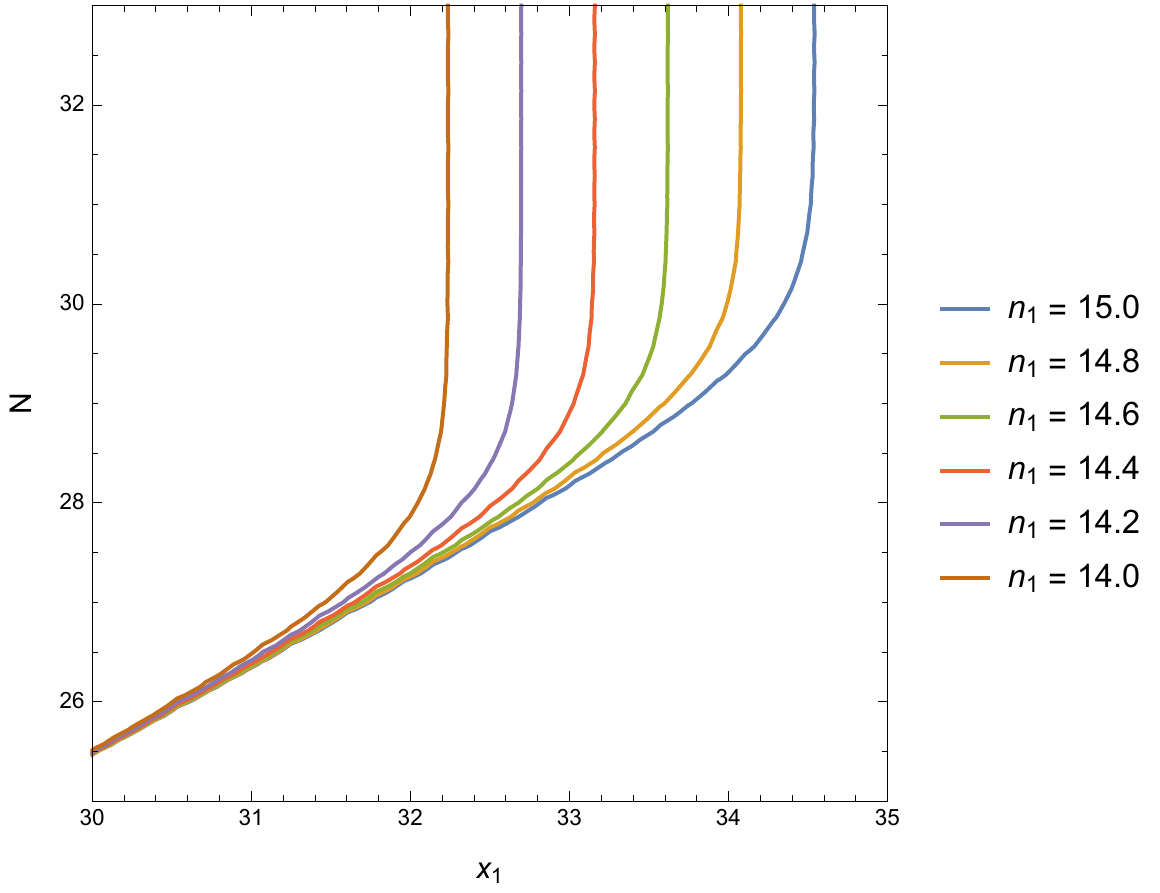}
\caption{\begin{footnotesize} Plot of $N$ versus $x_1$ for a few representative values of $n_1=14-15$, for positive brane cosmological constant. For small departure from the limiting value $x_1=n_1\textrm{ln}10$, the magnitude of $N$ falls sharply, leading to corresponding power law increase in the value of the cosmological constant $\omega_2^2=10^{-N}$. \end{footnotesize}}
\label{fig2}
\end{figure}

An interesting feature that generalizes readily from the singly warped counterpart is the variation of $N$ (defined here via $\omega_2^2=10^{-N}$) with $x_1$. Away from the divergence at the flat limit $x_1=n_1\textrm{ln}10$, the value of $N$ decreases sharply as shown in Figure \ref{fig2}, leading to rapid increase of $\omega_2^2$. Thus, the observed tiny value of the cosmological constant requires $x_1$ (hence $\alpha$) to be very close to the flat limit. Based on the relation between $\alpha$ and $\beta$ from \eqref{bsol11} and that between $\beta$ and $r_z$ from \eqref{bsol12}, this naturally explains the tuning of $r_z$ close to the Planck length when the bulk cosmological constant $\Lambda$ is of order $-M^6$. 
\\

\subsection{Dominant warping along $z$ ($\alpha<<\beta$)} \label{posomz}
Like its AdS counterpart, this regime admits a vanishingly small $\alpha\sim10^{-14}-10^{-12}$ and a unique $\beta\simeq12$ to generate the large mass hierarchy without any unnaturally large $R_y/r_z$ ratio. Consequently, the metric of the five-dimensional submanifold is left almost unwarped along $y$. 

In accordance with \eqref{V22}, the criterion for $V_2(z)$ to be positive is given by
\begin{equation}
\textrm{coth}\left(\textrm{ln}\dfrac{c_2}{\omega_2}-\alpha\pi\right)\leq0\:\:\implies\:\:\dfrac{\omega_2^2}{c_2^2}e^{x_1}\geq e^{-x_1}
\end{equation}
Owing to the positivity of the warp factor $a(\pi)$, this condition cannot be satisfied by any value of $\omega_2^2$ or $x_1$. So it is not possible to have a positive tension $(\pi,0)$ brane in this case either.

Similar to the AdS case, here also one can figure out how the clustering of the $\mathcal{O}$(TeV) mass scales depends on $\Omega$. The ratio between the scales of the maximally and near-maximally warped 3-branes remains equal to $a(\pi)$, which in this case yields the following result:
\begin{equation}
\dfrac{\textrm{Mass scale of maximally warped}\:(\pi,0)\:\textrm{brane}}{\textrm{Mass scale of near-maximally warped}\:(0,0)\:\textrm{brane}}=a(\pi)\:\approx\:1-\alpha\pi\sqrt{1+\omega_2^2}
\end{equation}
For $\omega_2=0$ this boils down to $(1-\alpha\pi)$ as expected. To arrive at non-trivial results, let us first recall that $\omega_2$ in this regime can be safely larger than unity as long as it satisfies $\omega_2/c_2<e^{-\alpha\pi}$ as meant to ensure the positivity of $a(\pi)$. This gives rise to three distinct possibilities: 
\begin{enumerate}
\item One may choose $\alpha\sim10^{-12}$ and $0<\omega_2\leq1$, for which the deviation of the ratio from its flat limit is too minute to be phenomenologically significant. 
\item As in the $\Omega<0$ case, the deviation in case of the previous possibility is more pronounced for larger values of $\alpha$, but they are all accompanied by extremely large $R_y/r_z$ ratios which render them unappealing from a physical standpoint.
\item Even for a conservative $\alpha\sim10^{-12}$ which results in no large $R_y/r_z$ hierarchy, choosing $\omega_2>>1$ (which still satisfies the inequality written above) can produce a significantly smaller value of $a(\pi)$ compared to its flat counterpart. 
\end{enumerate}
The first two possibilities are shared by the AdS case as seen earlier, while the third is unique to dS 3-branes and hinges on the possibility that $\omega_2$ can be larger than unity. In Table \ref{split2}, we list a few representative combinations of $\alpha$ and $\omega_2$ for both cases (2) and (3). The latter case implies $10^{-6}\lesssim\Omega R_y^2\lesssim10^{-2}$ for the values that have been chosen, but this magnitude can be lowered for lesser values of $\omega_2$ (i.e. at the cost of closer clustering). Note that the second and third columns are the same as those of Table \ref{split1}, as the same flat limit is recovered for both $\omega_1=0$ and $\omega_2=0$, and the functional dependence of the $R_y/r_z$ ratio on $\alpha$ and $\beta$ is the same in the $\Omega>0$ case as for $\Omega<0$. Furthermore, the clustering ratio $a(\pi)$ is smaller than its flat limit for \textit{any} non-zero $\omega_2$, i.e. the scales of the maximally and near-maximally warped branes are farther apart from each other when the branes are positively curved than when they are flat - which is the opposite of the corresponding AdS behaviour. 
\\ \\
The clustering ratio $a(\pi)$ can be readily related to the mass splitting ratio ($r$) of two fermions localized on the proximal branes, which is an interesting feature of multiply warped models. Following the method shown in \cite{fermmass}, it is easy to see that in our case $r=a(\pi)^{3/2}$. The fact that the degree of clustering, hence the proximity of fermion masses, increases with decreasing $\omega_2$ is interesting from a cosmological standpoint. First, let us consider the case of any two SM fermions located on the proximal $\mathcal{O}$(TeV) branes. Depending on the choice of SM fermions, the observed magnitude of $r$ can range from $10^{-2}$ (between the $172$ GeV top quark and the $4.2$ GeV bottom quark) to $10^{-12}$ (between the top quark and the $\sim1$ eV neutrino mass scale). Now, we firstly need $\alpha\lesssim10^{-10}$ to preclude any $R_y/r_z$ hierarchy larger than $10^4$. Even this threshold value leads to the requirement of a large $\omega_2$ of order at least $10^9$ if we wish to generate $r\sim0.01$ (smaller values of $r$ require even larger values of $\omega_2$ tuned sufficiently close to its allowed maximum value). Assuming $r_z\sim M^{-1}$ and hence $R_y\sim10^4M^{-1}$, one ends up with $\Omega/M^2\sim10^{-10}$ for such a high value of $\omega_2$ (it can easily be shown that $w_2\sim10^n$ roughly corresponds to $\Omega/M^2\sim10^{2n-30}$ in the small $\alpha$ regime), which is far too large compared to its observed value of about $10^{-120}$. At first glance, this rules out using $\omega_2$ as a regulator to consistently generate the presently observed mass hierarchy between any two distinct SM fermions on the clustered branes. However, in conjunction with other independent mechanisms (see for example \cite{cosmconst1,cosmconst2,cosmconst3,cosmconst4,cosmconst5,cosmconst6}) which attempt to resolve the discrepancy between the theoretically predicted large value of $\Omega$ and its observed small value, the cosmological constant could still be made to function as a tuning parameter for $r$. However, the need to invoke such external mechanisms renders the entire setup non-minimal in nature and less appealing.

Alternatively and perhaps more interestingly, instead of readily generating such a hierarchy, the model offers room for an intriguing hypothesis. Conventional wisdom suggests that a much larger cosmological constant may have prevailed during the very early stages of the universe, with the end of inflation marking its switchover to the present scale. As we have just seen, such a large value of $\Omega$ can produce $r\lesssim0.1$ without difficulty. On the other hand, when the inflaton has started decaying into the SM spectrum during the process of preheating, and the cosmological constant has not yet been reduced all the way down (say $\Omega/M^2\sim10^{-12}$), the number of degrees of freedom is naturally expected to be quite high. Among the particles produced under such conditions, one may hypothesize the existence of multiple species of fermions with nearly identical physical properties but masses separated by orders of $10^{-1}$ or $10^{-2}$. In our model, such a mass splitting can be interpreted as a result of the large value of $\Omega$, providing an unified framework for both ideas. By the end of the reheating period, when $\Omega$ has reached its currently observed value, this hierarchy becomes extremely close to unity (up to $\mathcal{O}(10^{-x})$ for $\alpha\sim10^{-x}$ with $x\sim\mathcal{O}(10)$), and the different species can effectively appear as a single species to low energy observers for the rest of cosmic history. Thus, an SM fermion observed today could actually be a family of distinct but nearly-identical fermions with an extremely close-knit mass spectrum which may be probed directly only at the GUT scale or beyond, whereas in the early universe they might have had a far more pronounced splitting induced by the larger cosmological constant. In the doubly warped setting, this works for only two fermions, while higher dimensional extensions allow the inclusion of more species due to a larger number of near-maximally warped 3-branes. A full treatment of this idea requires a study of the cosmological dynamics of the curved doubly-warped braneworld model, which falls outside our current scope. So we mention this here only as an interesting possibility offered by this model, and defer such a treatment to a future work.
\\
\begin{table}[ht]
\begin{center}
\begin{tabular}{|c|c|c|c|c|}
\hline
$\alpha$ & $a(\pi)$ in flat limit & $\sim R_y/r_z$ & $\omega_2$ & $a(\pi)$ for given $\omega_2$ \\
\hline\hline
\multirow{3}{*}{$10^{-6}$} & \multirow{3}{*}{$0.9999968584$} & \multirow{3}{*}{$\sim10^9$} & $0.3$ & $0.9999967201$ \\ & & & $0.6$ & $0.9999963363$ \\ & & & $0.9$ & $0.9999957734$ \\
\hline
\multirow{3}{*}{$10^{-4}$} & \multirow{3}{*}{$0.9996858901$} & \multirow{3}{*}{$\sim10^{11}$} & $0.3$ & $0.9996720574$ \\ & & & $0.6$ & $0.9996336798$ \\ & & & $0.9$ & $0.9995773913$ \\
\hline
\multirow{3}{*}{$10^{-2}$} & \multirow{3}{*}{$0.9690724263$} & \multirow{3}{*}{$\sim10^{13}$} & $0.3$ & $0.9676889351$ \\ & & & $0.6$ & $0.9638505427$ \\ & & & $0.9$ & $0.9582207615$ \\
\hline\hline\hline
\multirow{3}{*}{$10^{-14}$} & \multirow{3}{*}{$\sim1-10^{-14}$} & \multirow{3}{*}{$\sim10^1$} & $10^{11}$ & $0.9968643188$ \\ & & & $10^{12}$ & $0.9687500000$ \\ & & & $10^{13}$ & $0.6865234375$ \\
\hline
\multirow{3}{*}{$10^{-13}$} & \multirow{3}{*}{$\sim1-10^{-13}$} & \multirow{3}{*}{$\sim10^{2}$} & $10^{10}$ & $0.9968585968$ \\ & & & $10^{11}$ & $0.9685821533$ \\ & & & $10^{12}$ & $0.6858520508$ \\
\hline
\multirow{3}{*}{$10^{-12}$} & \multirow{3}{*}{$\sim1-10^{-12}$} & \multirow{3}{*}{$\sim10^{3}$} & $10^9$ & $0.9968584776$ \\ & & & $10^{10}$ & $0.9685850143$ \\ & & & $10^{11}$ & $0.6858444214$ \\
\hline
\end{tabular}
\end{center}
\vspace*{-2mm}
\caption{\begin{footnotesize} A few representative combinations of $\alpha$ and $\omega_2$ which are shown to considerably affect the clustering of the TeV-branes in the $\alpha<<\beta$ regime, as compared to the flat limit result. The first three sets of $\alpha$, corresponding to case (2) from Section \ref{posomz}, are accompanied by dangerously large radii ratios which introduce a new hierarchy in the model. The last three sets, corresponding to case (3), are free from this issue and make use of $\omega_2>>1$ allowed in the dS regime. For all the combinations, $\beta\simeq12$ has been fixed beforehand to obtain the desired amount of dominant warping along $z$. \end{footnotesize}}
\label{split2}
\end{table}

\section{Prospect of near-maximally warped visible brane} \label{nomax}
So far we have identified the maximally warped $(\pi,0)$ brane as the visible brane, based on the conservative assumption that the latter should have the least possible mass scale among all the existing 3-branes. As shown in the preceding sections, this identification allows a positive tension visible brane only for $\Omega<0$. Relaxing this assumption, we may identify the near-maximally warped $(\pi,\pi)$ brane as the visible brane instead, but only at the cost of allowing a brane with a slightly lower physical mass scale to exist in our system. In this section, we briefly focus on the implications of such a possibility for the visible brane tension, for each of the four cases considered so far.

\subsection{$\Omega<0$} \label{nomaxneg}
For $\alpha<<\beta$, the near-maximally warped brane lies at $(0,0)$. Its tension is equal to $V_1(0)$ from \eqref{V11}, which is positive by default for all values of $\omega_1$. Thus, its positivity does not require any fine tuning of $\omega_1$, unlike that of the $(\pi,0)$ brane. This aspect makes it preferable to choose the $(0,0)$ brane as the visible brane in this case, which of course comes at the cost of allowing a slightly lower mass-scale $(\pi,0)$ brane that can have negative tension in absence of such fine tuning. Also, to resolve the hierarchy problem at $(0,0)$, we must have $n_2\simeq16$ (corresponding to $\beta\simeq12$), as $a(0)=1$ contributes nothing to the warping along $y$. The upper bound on $|\Omega|R_y^2$ from Section \ref{negomz} holds good, and in fact, is allowed to be even smaller as $\omega_1$ does not need to be tuned close to $1$ anymore. 

For $\alpha>\beta$, the near-maximally warped brane is the $(\pi,\pi)$ brane. Its tension is equal to $V_2(\pi)+V_4(\pi)$, which is given explicitly by
\begin{equation} \label{vten1}
V_2(\pi)+V_4(\pi)=8M^2\sqrt{-\dfrac{\Lambda_6}{40}}\left[4\:\textrm{tanh}(x_2)+3\:\textrm{tanh}\left(\textrm{ln}\dfrac{\omega_1}{c_1}+x_1\right)\textrm{sech}(x_2)\right]
\end{equation}
For the second solution $x_1^{(2)}$ from \eqref{exsolexp2}, the coefficient term of $\textrm{sech}(x_2)$ approaches $+1$ as shown in \eqref{tanh}, which renders the tension of this brane positive. For the other solution $x_1^{(1)}$, it approaches $-1$ and the setup mimicks the flat limit. In that case, the tension can still be positive for any $x_2>x_{2r}$, where $x_{2r}$ is the root of the equation $4\textrm{tanh}(x_{2r})=3\textrm{sech}(x_{2r})$, given by $x_{2r}\:\approx\:0.694$ (or equivalently $\beta_r\:\approx\:0.221$). On the other hand, in order to steer clear of a large $R_y/r_z$ ratio, $x_2$ cannot be arbitrarily larger than this lower limit. For example, $x_2$ can comfortably range from $0.69$ to $2.50$ (say), where the latter is an upper limit assumed to be set by some modulus stabilization mechanism. The key takeaway is the fact that a near-maximally warped visible brane can have positive tension for both the solutions of $x_1$ that resolve the hierarchy problem, unlike only the second solution $x_1^{(2)}$ as in Section \ref{negomy}. 

\subsection{$\Omega>0$} \label{nomaxpos}
As before, for $\alpha<<\beta$, the tension of the near-maximally warped $(0,0)$ brane is given by $V_1(0)$ which is unambiguously positive. This marks a notable departure from Section \ref{posomz}, where it was impossible to have a positive tension visible brane. 

For $\alpha>\beta$, the near-maximally warped $(\pi,\pi)$ brane has tension
\begin{equation} \label{vten2}
V_2(\pi)+V_4(\pi)=8M^2\sqrt{-\dfrac{\Lambda_6}{40}}\left[4\:\textrm{tanh}(x_2)-3\:\textrm{coth}\left(\textrm{ln}\dfrac{c_2}{\omega_2}-x_1\right)\textrm{sech}(x_2)\right]
\end{equation}
Using \eqref{asol2} along with $a(\pi)=10^{-n_1}$ and exploiting the positivity of the warp factor, the condition for the positivity of the tension can be recast as
\begin{equation} \label{w2bound}
4\:\textrm{tanh}(x_2)-3\left(\sqrt{1+\omega_2^210^{2n_1}}\right)\textrm{sech}(x_2)>0\:\:\implies\:\:\omega_2<\dfrac{1}{3}\sqrt{16\:\textrm{sinh}^2(x_2)-9}\times10^{-n_1}
\end{equation}
This sets an upper bound to the magnitude of $\omega_2$. Firstly, for a real value of this upper bound, one needs $x_2\gtrsim0.694$. This is quite expectedly identical to the value of $x_{2r}$ calculated earlier, as in the $\Omega\to0$ limit \eqref{vten1} and \eqref{vten2} become identical, and any $x_2>x_{2r}$ can satisfy \eqref{w2bound} trivially and give a positive tension. But $x_2$ cannot be arbitrarily large as argued earlier, and needs to be bounded above by an $\mathcal{O}(1)$ upper limit as well. This, in turn, safely ensures that the scales of the maximally warped $(\pi,0)$ brane and $(\pi,\pi)$ are sufficiently close. Secondly, to resolve the hierarchy problem on $(\pi,\pi)$, we require $n_1=16$, as $b(\pi)=1$ gives zero warping along $z$. Together, these conditions furnish a small upper bound of roughly $\omega_2\lesssim10^{-15}$. Note that this is consistent with the $\omega_2/c_2<e^{-\alpha\pi}$ criterion necessary for the positivity of $a(\pi)$, which was observed earlier in Section \ref{posom}.

Thus, in the $\Omega>0$ regime, the tension of the near-maximally warped brane can be positive for both $\alpha<<\beta$ and $\alpha>\beta$, with the latter case additionally entailing a small upper bound on the magnitude of the brane cosmological constant as a requisite criterion. These features differ substantially from the results obtained in Sections \ref{posomy} and \ref{posomz}, and render the possibility of a near-maximally warped visible brane in this regime far more interesting. 

\section{Higher dimensional extensions} \label{highdim}
The analysis so far can be readily extended to spacetimes with $n$-fold warping over a series of nested $S^1/\mathbb{Z}_2$ orbifolds, equipped with $(n+4)$-dimensional metrics of the form
\begin{equation} \label{ndim}
ds^2=a_n(y_n)^2\left[a_{n-1}(y_{n-1})^2...\left[a_1(y_1)^2g_{\mu\nu}dx^\mu dx^\nu+R_1^2dy_1^2\right]+...+R_{n-1}^2dy_{n-1}^2\right]+R_n^2dy_n^2
\end{equation}
where $R_i$s are the orbifold radii, $y_i$s are the compact angular coordinates, and $a_i$s are the corresponding warp factors. Evidently, the metric elements $\tilde{g}_{AB}$ are of the forms:
\begin{equation}
\tilde{g}_{\mu\nu}=g_{\mu\nu}\prod_{i=1}^na_i(y_i)^2\:,\:\:\tilde{g}_{jj}=R_j^2\prod_{i=j+1}^na_i(y_i)^2\:,\:\:\tilde{g}_{nn}=R_n^2
\end{equation}
where $\tilde{g}_{jj}$ are the extra dimensional components. The bulk cosmological constant is denoted by $\Lambda_{n+4}$. The setup consists of $2n$ number of $p$-branes where $p=n+2$, with one pair of branes at every set of fixed points $y_i=\{0,\pi\}$. The intersections of these branes produce $2^n$ number of 3-branes at the vertices of the resulting ``brane-box" configuration. Since the effect of curvature enters principally through the first level of warping, we expect the geometric properties of these extended models to closely mimick those of the curved doubly warped scenario. We demonstrate the validity of this claim by first analyzing the triply warped case, i.e., $n=3$, for which the EFEs take the following forms (where primes denote derivatives wrt respective variables): 
\\ \\
\underline{$\mu\nu$-component}:
\\ \\
\begin{footnotesize}
\begin{equation} \label{7dmunu}
\begin{split}
& G_{\mu\nu}+g_{\mu\nu}\left(\dfrac{3a_1a_1''}{R_1^2}+\dfrac{3a_1'^2}{R_1^2}+\dfrac{4a_1^2a_2a_2''}{R_2^2}+\dfrac{6a_1^2a_2'^2}{R_2^2}+\dfrac{5a_1^2a_2^2a_3a_3''}{R_3^2}+\dfrac{10a_1^2a_2^2a_3'^2}{R_3^2}\right) \\
& = -\dfrac{\Lambda_7}{4M^5}a_1^2a_2^2a_3^2g_{\mu\nu}-\dfrac{a_1^2a_2a_3}{4M^5}g_{\mu\nu}\Bigg[\dfrac{1}{R_1}\Big\{V_1(y_2,y_3)\delta(y_1)+V_2(y_2,y_3)\delta(y_1-\pi)\Big\} \\
& \quad +\dfrac{a_2}{R_2}\Big\{V_3(y_1,y_3)\delta(y_2)+V_4(y_1,y_3)\delta(y_2-\pi)\Big\}+\dfrac{a_2a_3}{R_3}\Big\{V_5(y_1,y_2)\delta(y_3)+V_6(y_1,y_2)\delta(y_3-\pi)\Big\}\Bigg]
\end{split}
\end{equation}
\end{footnotesize}
\\ \\
\underline{$y_1y_1$-component}:
\\ \\
\begin{footnotesize}
\begin{equation} \label{7d55}
\begin{split}
& 2M^5\left[\mathcal{R}-\dfrac{12a_1'^2}{R_1^2}-\dfrac{4a_1^2}{R_2^2}\left(2a_2a_2''+3a_2'^2\right)-\dfrac{10a_1^2a_2^2}{R_3^2}\left(a_3a_3''+2a_3'^2\right)\right] \\
& =a_1^2a_2^2a_3^2\left[\Lambda_7+\dfrac{V_3(y_1,y_3)\delta(y_2)+V_4(y_1,y_3)\delta(y_2-\pi)}{a_3R_2}+\dfrac{V_5(y_1,y_2)\delta(y_3)+V_6(y_1,y_2)\delta(y_3-\pi)}{R_3}\right]
\end{split}
\end{equation}
\end{footnotesize}
\\ \\
\underline{$y_2y_2$-component}:
\\ \\
\begin{footnotesize}
\begin{equation} \label{7d66}
\begin{split}
& 2M^5\left[\mathcal{R}-\dfrac{20a_1^2a_2'^2}{R_2^2}-\dfrac{4}{R_1^2}\left(2a_1a_1''+3a_1'^2\right)-\dfrac{10a_1^2a_2^2}{R_3^2}\left(a_3a_3''+2a_3'^2\right)\right] \\
& =a_1^2a_2^2a_3^2\left[\Lambda_7+\dfrac{V_1(y_2,y_3)\delta(y_1)+V_2(y_2,y_3)\delta(y_1-\pi)}{a_2a_3R_1}+\dfrac{V_5(y_1,y_2)\delta(y_3)+V_6(y_1,y_2)\delta(y_3-\pi)}{R_3}\right]
\end{split}
\end{equation}
\end{footnotesize}
\\ \\
\underline{$y_3y_3$-component}:
\\ \\
\begin{footnotesize}
\begin{equation} \label{7d77}
\begin{split}
& 2M^5\left[\mathcal{R}-\dfrac{30a_1^2a_2^2a_3'^2}{R_3^2}-\dfrac{4}{R_1^2}\left(2a_1a_1''+3a_1'^2\right)-\dfrac{10a_1^2}{R_2^2}\left(a_2a_2''+2a_2'^2\right)\right] \\
& =a_1^2a_2^2a_3^2\left[\Lambda_7+\dfrac{V_1(y_2,y_3)\delta(y_1)+V_2(y_2,y_3)\delta(y_1-\pi)}{a_2a_3R_1}+\dfrac{V_3(y_1,y_3)\delta(y_2)+V_4(y_1,y_3)\delta(y_2-\pi)}{R_3}\right]
\end{split}
\end{equation}
\end{footnotesize}
\\ \\
In the bulk, we can separate the $\mu\nu$ and extra dimensional variables from \eqref{7dmunu} and write $G_{\mu\nu}+\Omega g_{\mu\nu}=0$, where the effective cosmological constant $(\Omega)$ is defined as 
\begin{equation} \label{7domega}
\dfrac{3}{R_1^2}\left(a_1a_1''+a_1'^2\right)+\dfrac{2a_1^2}{R_2^2}\left(2a_2a_2''+3a_2'^2\right)+\dfrac{5a_1^2a_2^2}{R_3^2}\left(a_3a_3''+2a_3'^2\right)+\dfrac{\Lambda_7}{4M^5}a_1^2a_2^2a_3^2=\Omega
\end{equation}
Using $\mathcal{R}=4\Omega$ from the 4D equation and plugging \eqref{7domega} for $\Omega$ in \eqref{7d55}, the bulk cosmological constant can be expressed as
\begin{equation} \label{7dbulk}
\left(\dfrac{a_1^2a_2^2a_3^2}{4M^5}\right)\Lambda_7=-\dfrac{6a_1a_1''}{R_1^2}-\dfrac{2a_1^2}{R_2^2}\left(2a_2a_2''+3a_2'^2\right)-\dfrac{5a_1^2a_2^2}{R_3^2}\left(a_3a_3''+2a_3'^2\right)
\end{equation}
The steps to uncouple the warp factors are similar to the doubly warped case: (i) Plug \eqref{7dbulk} into \eqref{7d77} to eliminate $\Lambda_7$. (ii) Compare \eqref{7d55} and \eqref{7d77} to express $a_3a_3''-a_3'^2$ in terms of $a_1$ and $a_2$ $-$ a result which when inserted into the previous equation gives the first uncoupled ODE for $a_1(y_1)$. (iii) Now compare \eqref{7d66} and \eqref{7d77} to eliminate $a_3$ in terms of $a_2$ alone, and insert into the same equation to obtain the second uncoupled ODE for $a_2(y_2)$. (iv) In the eliminating relation found and used in the previous step, rearrange terms to construct the final ODE for $a_3(y_3)$. The equations thus obtained have remarkably simple forms.
\begin{equation} \label{eqa1}
a_1'^2-a_1a_1''=\dfrac{\Omega R_1^2}{3}\:\:\:;\:\:\:a_1''-\alpha_1^2a_1=0
\end{equation}
\begin{equation} \label{eqa2}
a_2'^2-a_2a_2''=-\dfrac{\alpha_1^2R_2^2}{R_1^2}\:\:\:;\:\:\:a_2''-\alpha_2^2a_2=0
\end{equation}
\begin{equation} \label{eqa3}
a_3'^2-a_3a_3''=-\dfrac{\alpha_2^2R_3^2}{R_2^2}
\end{equation}
where $\alpha_1$ and $\alpha_2$ are two dimensionless constants of separation. With appropriate normalization, the final solutions of the warp factors are given by:
\begin{center}
\begin{tabular}{|c|c|}
\hline
$\Omega<0$ & $\Omega>0$ \\
\hline\hline
$\:$ & $\:$ \\
$\quad   a_1(y_1)=\omega_1\textrm{cosh}\left(\textrm{ln}\dfrac{\omega_1}{c_1}+\alpha_1|y_1|\right)   \quad$ & $\quad   a_1(y_1)=\omega_2\:\textrm{sinh}\left(\textrm{ln}\dfrac{c_2}{\omega_2}-\alpha_1|y_1|\right)   \quad$ \\
$\omega_1^2=-\dfrac{\Omega R_1^2}{3\alpha_1^2}\:\:,\:\:\:c_1=1+\sqrt{1-\omega_1^2}$ & $\omega_2^2=\dfrac{\Omega R_1^2}{3\alpha_1^2}\:\:,\:\:\:c_2=1+\sqrt{1+\omega_2^2}$ \\
$\:$ & $\:$ \\
\hline
$\:$ & $\:$ \\
$\quad   a_2(y_2)=\dfrac{\textrm{cosh}(\alpha_2y_2)}{\textrm{cosh}(\alpha_2\pi)}\:,\:\:\alpha_1=\dfrac{R_1\alpha_2}{R_2\textrm{cosh}(\alpha_2\pi)}   \quad$ & $\quad   a_2(y_2)=\dfrac{\textrm{cosh}(\alpha_2y_2)}{\textrm{cosh}(\alpha_2\pi)}\:,\:\:\alpha_1=\dfrac{R_1\alpha_2}{R_2\textrm{cosh}(\alpha_2\pi)}   \quad$ \\
$\:$ & $\:$ \\
\hline
$\:$ & $\:$ \\
$\quad   a_3(y_3)=\dfrac{\textrm{cosh}(\alpha_3y_3)}{\textrm{cosh}(\alpha_3\pi)}\:,\:\:\alpha_2=\dfrac{R_2\alpha_3}{R_3\textrm{cosh}(\alpha_3\pi)}   \quad$ & $\quad   a_3(y_3)=\dfrac{\textrm{cosh}(\alpha_3y_3)}{\textrm{cosh}(\alpha_3\pi)}\:,\:\:\alpha_2=\dfrac{R_2\alpha_3}{R_3\textrm{cosh}(\alpha_3\pi)}   \quad$ \\
$\:$ & $\:$ \\
\hline
\end{tabular}
\end{center}
\vspace*{2mm}
where $\alpha_3/R_3=\sqrt{-\Lambda_7/(60M^5)}$ analogous to \eqref{bsol12}. Once again, it is clear that the effect of $\Omega$ appears explicitly only in $a_1(y_1)$. The equations (and solutions) of the other warp factors appear thereafter in a hierarchical manner, with nested functional relations among the warping parameters and the orbifold radii. Equipped with these solutions, the brane tensions can be found by integrating \eqref{7dmunu} across each fixed point. The non-zero tensions are:
\begin{equation}
\Omega<0\:\implies\:
\begin{dcases}
V_1=24M^{\frac{5}{2}}\sqrt{-\dfrac{\Lambda_7}{60}\left(1-\omega_1^2\right)}\:\textrm{sech}(\alpha_2y_2)\:\textrm{sech}(\alpha_3y_3) \\
V_2=24M^{\frac{5}{2}}\sqrt{-\dfrac{\Lambda_7}{60}}\:\textrm{tanh}\left(\textrm{ln}\dfrac{\omega_1}{c_1}+\alpha_1\pi\right)\textrm{sech}(\alpha_2y_2)\:\textrm{sech}(\alpha_3y_3)
\end{dcases}
\end{equation}
\begin{equation}
\Omega>0\:\implies\:
\begin{dcases}
V_1=24M^{\frac{5}{2}}\sqrt{-\dfrac{\Lambda_7}{60}\left(1+\omega_2^2\right)}\:\textrm{sech}(\alpha_2y_2)\:\textrm{sech}(\alpha_3y_3) \\
V_2=-24M^{\frac{5}{2}}\sqrt{-\dfrac{\Lambda_7}{60}}\:\textrm{coth}\left(\textrm{ln}\dfrac{c_2}{\omega_2}-\alpha_1\pi\right)\textrm{sech}(\alpha_2y_2)\:\textrm{sech}(\alpha_3y_3)
\end{dcases}
\end{equation}
\begin{equation}
V_4=32M^{\frac{5}{2}}\sqrt{-\dfrac{\Lambda_7}{60}}\:\textrm{tanh}(\alpha_2\pi)\:\textrm{sech}(\alpha_3y_3)\:,\:\:V_6=40M^{\frac{5}{2}}\sqrt{-\dfrac{\Lambda_7}{60}}\:\textrm{tanh}(\alpha_3\pi)
\end{equation}
Note that only the compact coordinates of higher orbifolds contribute to the coordinate dependence of each brane tension, which is a consequence of nested warping. 

\begin{figure}[h]
\centering
\begin{minipage}[h]{0.45\textwidth}
\centering
\includegraphics[width=0.95\textwidth,height=0.85\textwidth]{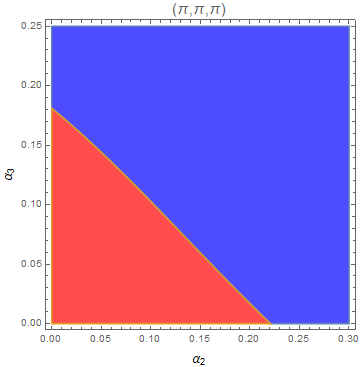}
\label{fig41}
\end{minipage}
\begin{minipage}[h]{0.45\textwidth}
\centering
\includegraphics[width=0.95\textwidth,height=0.85\textwidth]{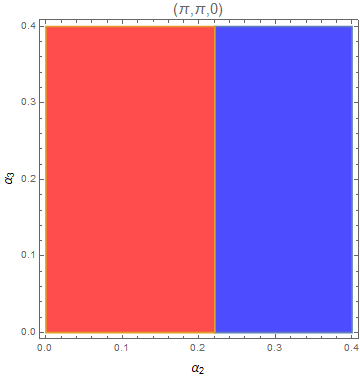}
\label{fig42}
\end{minipage}
\begin{minipage}[h]{0.45\textwidth}
\centering
\includegraphics[width=0.95\textwidth,height=0.85\textwidth]{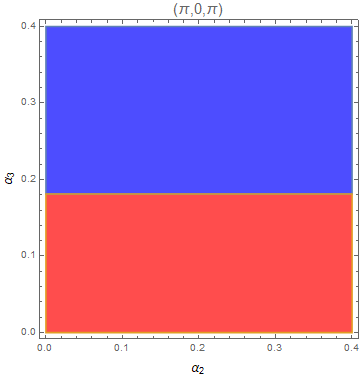}
\label{fig43}
\end{minipage}
\begin{minipage}[h]{0.45\textwidth}
\centering
\includegraphics[width=0.95\textwidth,height=0.85\textwidth]{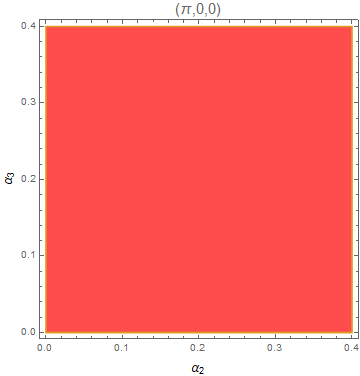}
\label{fig44}
\end{minipage}
\caption{\begin{footnotesize} For $\Omega<0$ and $\alpha_1\sim10$, partition of the $\{\alpha_2,\alpha_3\}$ parameter space of the curved triply warped model corresponding to the signs of the tension of each near-maximally warped 3-brane, for the first solution $\alpha_1=\alpha_1^{(1)}$ from \eqref{exsolexp1}. The red (bounded) and blue (unbounded) regions in the first three cases correspond to negative and positive tensions respectively. Note that for the maximally warped $(\pi,0,0)$ brane, there is no blue region for $\alpha_1=\alpha_1^{(1)}$ as the tension is identically negative, whereas for $\alpha_1=\alpha_1^{(2)}$ the reverse situation arises. \end{footnotesize}}
\label{figure4}
\end{figure}

In order to avoid a large hierarchy among the radii, the regimes allowed are either $\alpha_1\sim10$ alongside $\alpha_2\sim\alpha_3\sim0.1$, or any one of $\alpha_2$ or $\alpha_3$ close to 12 with the other two being vanishingly small $(\sim10^{-15})$. Regardless of the choice of regime, maximum possible warping always occurs at $\{y_1=\pi,\:y_2=0,\:y_3=0\}$. Identifying it with the visible brane leads to visible brane tension equal to $V_2(\pi)$, which, owing to its form, gives nothing quantitatively different from Sections \ref{negom} and \ref{posom}. The case of a non-maximally warped visible brane is arguably more interesting, for which one can proceed as in Section \ref{nomax} and analyze various possible configurations. As an example, consider the case $\Omega<0$ and $\alpha_1\sim10$, for which $(\pi,\pi,\pi)$ is a near-maximally warped brane with tension $V_2(\pi,\pi)+V_4(\pi)+V_6$. Recall that the hierarchy problem can be resolved for two distinct values $\alpha_1^{(1)}$ and $\alpha_1^{(2)}$ (refer back to \eqref{exsolexp1}), which render the $\textrm{tanh}$ term in $V_2$ respectively $-1$ and $+1$ . The latter identically makes the tension positive for all $\alpha_2$ and $\alpha_3$, whereas for the former, the positivity condition can be written explicitly as
\\
$$-3\:\textrm{sech}(\alpha_2\pi)\:\textrm{sech}(\alpha_3\pi)+4\:\textrm{tanh}(\alpha_2\pi)\:\textrm{sech}(\alpha_3\pi)+5\:\textrm{tanh}(\alpha_3\pi)>0$$
\begin{equation} \label{ineq}
\implies\:\:4\:\textrm{tanh}(\alpha_2\pi)-3\:\textrm{sech}(\alpha_2\pi)>-5\:\textrm{sinh}(\alpha_3\pi)
\end{equation}
\\
Clearly, for any given $\alpha_3>0$, the lower bound on $\alpha_2$ is smaller than the value of $0.221$ obtained in Section \ref{nomaxneg}. In fact, it can be readily checked that for $\alpha_3\gtrsim0.182$, any $\alpha_2>0$ can satisfy \eqref{ineq}. The triply warped model thus allows greater flexibility over the doubly warped one in rendering the tension positive. The cases of the other near-maximally warped branes can be analyzed individually in a similar fashion, thereby opening up different regions of parameter space consistent with a positive tension visible brane. In Figure \ref{figure4}, we show the allowed regions in the $\{\alpha_2,\alpha_3\}$ parameter space leading to positive tensions of each of the four near-maximally warped branes for $\Omega<0$ and $\alpha_1\sim10$, corresponding to the first solution $\alpha_1^{(1)}$ from \eqref{exsolexp1} that resolves the gauge hierarchy problem. 
\\
\\
The generalization to $n$ extra dimensions for the metric in \eqref{ndim} is straightforward at this point, for which the governing set of equations turns out to be
\begin{equation} \label{7deq1}
a_1'^2-a_1a_1''=\dfrac{\Omega R_1^2}{3}\:\:,\:\:\:a_j''-\alpha_j^2a_j=0
\end{equation}
\begin{equation} \label{7deq2}
a_{j+1}'^2-a_{j+1}a_{j+1}''=-\dfrac{\alpha_j^2R_{j+1}^2}{R_j^2}
\end{equation}
where $j$ ranges from $1$ to $n-1$. Depending on the sign of $\Omega$, the corresponding solutions of $a_1(y_1)$ are identical to the doubly (and triply) warped results. All the subsequent warp factors are independent of $\Omega$ and take the forms
\begin{equation} \label{7deq3}
a_j(y_j)=\dfrac{\textrm{cosh}(\alpha_jy_j)}{\textrm{cosh}(\alpha_j\pi)}\:\:,\:\:\:\alpha_{j-1}=\dfrac{R_{j-1}\alpha_j}{R_j\textrm{cosh}(\alpha_j\pi)}
\end{equation}
for $j=2$ to $n$, with $\alpha_n/R_n=\sqrt{-\Lambda_{n+4}/(20nM^{n+2})}$. The odd-labeled $p$-branes (with $p=n+2$) located at each $y_j=0$ for $j=2$ to $n$ have vanishing tension $V_{2j-1}=0$, while the remaining branes at $y_1=\{0,\pi\}$ and $y_j=\pi$ have tensions as follows:
\begin{equation} \label{ndimvneg}
\Omega<0\:\implies\:
\begin{dcases}
\:V_1=24M^{\frac{n+2}{2}}\sqrt{-\dfrac{\Lambda_{n+4}}{20n}\left(1-\omega_1^2\right)}\prod_{i=2}^n\:\textrm{sech}(\alpha_iy_i) \\
\:V_2=24M^{\frac{n+2}{2}}\sqrt{-\dfrac{\Lambda_{n+4}}{20n}}\:\textrm{tanh}\left(\textrm{ln}\dfrac{\omega_1}{c_1}+\alpha_1\pi\right)\prod_{i=2}^n\:\textrm{sech}(\alpha_iy_i)
\end{dcases}
\end{equation}
\begin{equation} \label{ndimvpos}
\Omega>0\:\implies\:
\begin{dcases}
\:V_1=24M^{\frac{n+2}{2}}\sqrt{-\dfrac{\Lambda_{n+4}}{20n}\left(1+\omega_2^2\right)}\prod_{i=2}^n\:\textrm{sech}(\alpha_iy_i) \\
\:V_2=-24M^{\frac{n+2}{2}}\sqrt{-\dfrac{\Lambda_{n+4}}{20n}}\:\textrm{coth}\left(\textrm{ln}\dfrac{c_2}{\omega_2}-\alpha_1\pi\right)\prod_{i=2}^n\:\textrm{sech}(\alpha_iy_i)
\end{dcases}
\end{equation}
\begin{equation} \label{ndimvhigher}
\:V_{2j}=8(j+2)M^{\frac{n+2}{2}}\sqrt{-\dfrac{\Lambda_{n+4}}{20n}}\:\textrm{tanh}(\alpha_j\pi)\prod_{i=j+1}^n\:\textrm{sech}(\alpha_iy_i) 
\end{equation}
where $\omega_1$, $c_1$, $\omega_2$, and $c_2$ are as defined earlier. To avoid a large hierarchy among the $R_j$s, the allowed regimes are either $\alpha_1\sim12$ with $\alpha_j\sim0.1$ for all $j=2$ to $n$, or any particular $\alpha_j\sim12$ with all the others (including $\alpha_1$) vanishingly small ($\sim10^{-15}$). Maximum warping occurs uniquely at $(\pi,0,...,0)$, with $2^{n-1}$ number of near-maximally warped 3-branes (with a closely spaced cluster of mass scales) arising in each of the allowed regimes. Proceeding as earlier, a case-by-case analysis of the tension of each near-maximal brane can be done for any given $n$. For example, in the $\Omega<0$ and $\alpha_1\sim10$ regime with $\alpha_1=\alpha_1^{(1)}$, there exists a non-trivial region in the $(n-1)$-dimensional $\{\alpha_2,...\:,\alpha_n\}$ parameter space for which the tension of the $(\pi,\pi,...\:,\pi)$ brane can be rendered positive (analogous to the top-left plot from Figure \ref{figure4}). Using \eqref{ndimvneg} and \eqref{ndimvhigher}, one can deduce that this region corresponds to the following inequality:
\begin{equation}
-3\prod_{i=2}^n\:\textrm{sech}(\alpha_i\pi)+\sum_{j=2}^n(j+2)\:\textrm{tanh}(\alpha_j\pi)\left(\prod_{i=j+1}^n\textrm{sech}(\alpha_i\pi)\right)>0
\end{equation}
Constraints for the other branes can be obtained similarly. Since $V_{2j-1}=0$ for all odd $j>1$, note that the inequality for a 3-brane produced by the intersection of any $y_j=0$ $p$-brane will be independent of $\alpha_j$. While in principle unbounded above, such regions in the parameter space cannot be arbitrarily large if we are to avoid - as noted earlier - an unnaturally large hierarchy among the various $R_i$s. An upper bound might be established, for instance, via some higher-dimensional analogue of the modulus stabilization mechanism explored in \cite{gravstab3}, a rigorous discussion of which lies beyond the scope of this work. This is an interesting question that we plan to address separately in near future. 
\\ \\
As for the dependence of the clustering of scales between the maximally warped and any near-maximally warped brane on the induced cosmological constant, the results for an arbitrary $n$-fold warped metric should not differ appreciably from the six dimensional case as $\Omega$ makes its appearance solely through $a_1(y_1)$ and in the same functional form. So the values of $a_1(\pi)$ from Tables \ref{split1} and \ref{split2} continue to hold, with other relevant $a_j(0)$ terms (independent of $\Omega$) appearing multiplicatively and providing subleading corrections to the overall scale ratio. 

\section{Conclusion}

In summary, we have generalized six and higher dimensional braneworld models with nested warping to include the effects of a non-zero cosmological constant $(\Omega)$ induced on the corner 3-branes, which results in global brane curvature.  We have first analyzed the doubly warped case in detail, for which the solutions of the warp factors from the Einstein field equations permit both negative and positive values of $\Omega$, corresponding respectively to AdS-Schwarzschild and dS-Schwarzschild geometries for the 3-branes. An important consequence of nested warping is that the effect of $\Omega$ appears explicitly only in the warp factor associated with the first orbifold. On the other hand, simultaneous large warping along both orbifolds is forbidden if the radii $R_y$ and $r_z$ of the compact spaces are of close magnitudes, which is a natural assumption to make. This is a salient feature of nested warped braneworld models that is seen to hold for both flat and curved scenarios. Together, these two properties result in $2\times2=4$ distinct possibilites which we have analyzed in detail, for which we have first assumed the maximally warped $(y=\pi,z=0)$ brane to be the visible brane accommodating our four-dimensional universe. 
\begin{itemize}
\item For $\Omega<0$, dominant warping along the first orbifold ($y$) imposes a small upper bound of order $10^{-32}$ on the value of $|\Omega|R_y^2$, which is interesting from a cosmological perspective. For values of the cosmological constant sufficiently smaller than this allowed maximum, the gauge hierarchy problem can be resolved for two distinct values of the warp factor $\alpha$ associated with this orbifold (as opposed to a unique value in the flat case). One of these solutions effectively reproduces the flat result, while the other one is distinct and non-trivial. Unlike the former, the latter can render the tension of the maximally warped 3-brane positive. On the other hand, dominant warping along the second orbifold ($z$) requires an extreme fine tuning of the cosmological constant if one wishes to simultaneously avoid a large $R_y/r_z$ hierarchy and have a positive tension $(\pi,0)$ brane. One also observes that the physical mass scales of the maximally warped $(0,0)$ brane and the near-maximally warped $(\pi,0)$ brane are clustered more closely in this regime than in the case of flat branes. If one strictly chooses $R_y/r_z\lesssim10^2$, this difference is imperceptibly small and unlikely to play any significant phenomenological role.
\item For $\Omega>0$, positivity of the warp factor along $y$ imposes a similar upper bound to the magnitude of $\Omega$. Assuming dominant warping along $y$ further renders it quite small. But unlike the AdS case, there is only one solution of $\alpha$ that can generate the desired large mass hierarchy on the visible brane. The magnitude of $\Omega$ rises exponentially for small deviations of the warping parameter from its flat value, thereby justifying a flat braneworld approximation from a physical perspective. Furthermore, the tension of the $(\pi,0)$ brane cannot be rendered positive for any set of values of the model parameters. Dominant warping along $z$ cannot provide a positive tension visible brane either. As for the clustering of the $(0,0)$ and $(\pi,0)$ branes' scales, they are found to be comparatively farther apart than their flat limit counterparts - which is opposite of the analogous AdS behaviour. Moreover, unlike the AdS case, large allowed values of the cosmological constant (e.g. $10^{-6}\lesssim\Omega R_y^2\lesssim10^{-2}$) can significantly modify the clustering ratio without jeopardizing a conservative $R_y/r_z$ ratio. This, however, is inconsistent with the much smaller value of $\Omega$ measured today, and prevents explaining the currently observed SM fermion mass hierarchy (along the lines of \cite{doubwarp} and \cite{fermmass}) by using the induced cosmological constant as a regulator, unless some separate mechanism (e.g. \cite{cosmconst1,cosmconst2,cosmconst3,cosmconst4,cosmconst5,cosmconst6}) to resolve the discrepancy between the theoretically predicted and observed values of $\Omega$ is invoked. Alternatively, one might hypothesize that end-of-inflation (p)reheating could have naturally led to the production of families of nearly-identical fermions each with such closely-spaced mass spectra as associated with a large cosmological constant (consistent with inflationary dynamics). Thereafter, the lowered value of $\Omega$ could have greatly reduced this mass hierarchy and rendered it extremely close to unity (up to $\mathcal{O}(10^{-12})$ or even smaller for conservative $R_y/r_z$ ratios), practically removing the distinction between different species to any low energy observer throughout the rest of cosmic history. This implies that any SM fermion observed today could actually be a tuple of two (for the doubly warped model) or more (for higher dimensional extensions) nearly-identical, fundamental fermions with extremely close masses that might only be distinguished at the GUT scale or higher. From a theoretical standpoint, this provides a unified framework which potentially accommodates both the natural consideration of additional degrees of freedom in the post-inflationary period (besides explaining why they are no longer distinctly detectable today) and the large cosmological constant briefly dominant during that period. In order to fully address this question however, one needs to study the cosmological aspects of this braneworld model in detail. This falls outside the scope of the present work, and we plan to address it in a future one.
\end{itemize}
A few of these features strikingly resemble those of the curved singly warped scenario analyzed in \cite{5dcurved1}. As mentioned earlier, this is but a consequence of secondary warping of the already-warped 5D metric from \cite{rs1} along a higher $S^1/\mathbb{Z}_2$ orbifold. Some further novelties concerning the visible brane tension can emerge by identifying, instead of the conventionally chosen maximally warped brane, the near-maximally warped 3-brane with our 4D universe. Recall that such a choice is possible only in nested braneworld models with at least two levels of warping. In the $\Omega<0$ regime, dominant warping along $z$ gives the near-maximal $(0,0)$ brane a positive tension by default. Dominant warping along $y$, on the other hand, can potentially render the tension of the $(\pi,\pi)$ brane positive for both solutions of $\alpha$ that generate the desired large mass hierarchy, as opposed to only the second solution which works for the maximally warped brane. In the opposite $\Omega>0$ regime, dominant warping along $z$ similarly results in a tension which is always positive. Dominant warping along $y$ can do the same as long as there is a small upper threshold to the value of $\Omega$. As readily apparent, these features differ significantly from what one obtains for a maximally warped visible brane. 
\\ \\
The curved doubly warped model can be readily generalized to seven and higher dimensions with a series of nested warpings over successive orbifolds. The equations for the warp factors, together with the relations among the warping parameters and the orbifold radii, follow a definite hierarchical pattern as seen from \eqref{7deq1}$-$\eqref{7deq3}. Since the effect of the induced cosmological constant enters only through the first level of warping, the dependence of scale-clustering between the maximal and any near-maximally warped brane on $\Omega$ remains virtually identical to the 6D case. Moreover, identifying near-maximal TeV-branes (instead of the conventionally chosen maximal one) as the visible brane can potentially open up non-trivial regions in the warping parameter space that can render the corresponding brane tension positive. 
\\ \\
Based on the results derived in this work, one can explore various phenomenological aspects of multiply warped non-flat braneworld models. The issue of moduli stabilization in curved higher dimensional scenarios is a particularly interesting avenue. Previously, the authors of the present work have demonstrated stabilization of the two moduli of the 6D flat warped braneworld model via a straightforward extension of the Goldberger-Wise mechanism \cite{modstab6d}. While a non-flat generalization of such a bulk field approach is expected to work well, it is tempting to check if modified gravity effects and/or non-zero brane curvature alone can produce a stabilizing potential without the need for any external field. This can indeed be realized in the non-flat 5D model, where the resulting radion has interesting phenomenological properties and cosmological implications \cite{gravstab2,gravstab5,gravstab3}. If such effective potentials governed solely by the gravitational sector can be shown to exist in multiply warped geometries as well, the phenomenology of the associated 4D scalar radions and their cosmological dynamics (e.g. their role in driving multi-field inflation) would be an area of considerable interest - one which we plan to explore in a future project. The inclusion of higher curvature terms in the gravitational action at sufficiently high energy scales (motivated here by the large bulk cosmological constant $\Lambda_6\sim-M^6$) is also likely to result in non-trivial modifications to the model. The effective dynamics of bulk matter and gauge fields against a curved warped brane background is another important area which needs to be focused on in greater detail. The issue of the mass discrepancy of the Standard Model $W$-boson \cite{wmass}, for example, falls within the purview of this domain. These questions lie beyond our current scope as well, and we plan to address them separately in future works. 

\section*{Acknowledgments}
Research work of AB is supported by the CSIR Junior Research Fellowship, Government of India.

\end{document}